\documentclass[apj, letterpaper]{emulateapj}
\usepackage{lscape}
  
\newcommand{\erg}{\mbox{$\rm\,erg$}}
\newcommand{\kev}{\mbox{$\rm\,keV$}}
\newcommand{\cm}{\mbox{$\rm\,cm$÷}}

\newcommand{\yr}{\mbox{$\rm\,yr$}}
\newcommand{\s}{\mbox{$\rm\,s$}}

\newcommand{\kpc}{\mbox{$\rm\,kpc$}}

\newcommand{\beq}{\begin{equation}}
\newcommand{\eeq}{\end{equation}}

\shorttitle{The Nature of X-ray Cavities}
\shortauthors{Diehl, Li, Fryer \& Rafferty}

\usepackage{natbib}
\newcommand{\bve}{\begin{verbatim}}
\newcommand{\eve}{\end{verbatim}}
\newcommand{\degree}{^{\rm o}}

\begin{document}

\title{Constraining the Nature of X-ray Cavities in Clusters and Galaxies}

\author{Steven Diehl and Hui Li}
\affil{Theoretical Astrophysics Group (T-6), Los Alamos National Laboratory, P.O. Box 1663, Los Alamos, NM 87545, USA}
\email{diehl@lanl.gov, hli@lanl.gov}

\author{Christopher L. Fryer}
\affil{Computational Physics and Methods Group (CCS-2), Los Alamos National Laboratory, P.O. Box 1663, Los Alamos, NM 87545, USA}
\and

\author{David Rafferty}
\affil{Astrophysical Institute, Department of Physics \& Astronomy, Ohio University, Athens, OH 45701, USA}
\affil{Department of Astronomy and Astrophysics, The Pennsylvania State
University, 525 Davey Laboratory, University Park, PA 16802, USA}

\begin{abstract}
We present results from an extensive survey of 64 cavities in the X-ray halos of clusters, groups and normal elliptical galaxies. We show that the evolution of the size of the cavities as they rise in the X-ray atmosphere is inconsistent with the standard model of adiabatic expansion of purely hydrodynamic models. We also note that the majority of the observed bubbles should have already been shredded apart by Rayleigh-Taylor and Richtmyer-Meshkov instabilities if they were of purely hydrodynamic nature. Instead we find that the data agrees much better with a model where the cavities are magnetically dominated and inflated by a current-dominated magneto-hydrodynamic jet model, recently developed by \citet{LiMHD1} and \citet{LiMHD2}. We conduct complex Monte-Carlo simulations of the cavity detection process including incompleteness effects to reproduce the cavity sample's characteristics. We find that the current-dominated model agrees within $1\sigma$, whereas the other models can be excluded at $>5\sigma$ confidence. To bring hydrodynamic models into better agreement, cavities would have to be continuously inflated. However, these assessments are dependent on our correct understanding of the detectability of cavities in X-ray atmospheres, and will await confirmation when automated cavity detection tools become available in the future. Our results have considerable impact on the energy budget associated with active galactic nucleus feedback.
\end{abstract}

\keywords{cooling flows --- X-rays: ISM --- X-rays: galaxies: clusters --- galaxies: clusters: general --- instabilities --- MHD }

%---------------------------------------------------------------------------------

\section{Introduction}\label{s.introduction}

One of the current most active areas of research in X-ray astronomy is aimed at solving the problem of feedback in elliptical galaxies, groups and clusters of galaxies. All of these systems suffer from the same unsolved problem that the gas in their cores has cooling times significantly shorter than their age, suggesting that the gas should already have cooled and formed stars \citep{McNamaraAGNReview}. This problem, historically referred to as the ``cooling flow problem'', points toward the existence of a heat source that is preventing the gas from cooling effectively in the center. Due to a lack of better understanding, this effect is generally referred to with the generic term ``feedback''.

Many possible feedback mechanisms have been proposed in recent years. The most basic idea is the conversion of gravitational energy into thermal energy, as gas falls into the center towards a deeper gravitational potential. However, this mechanism has been shown to only be significant in elliptical galaxies with a rather steep gravitational potential, such as NGC~6482 \citep{PonmanNGC6482}, and does not solve the cooling flow problem.

Another heat source that has often been suggested is feedback from supernovae. If the gas cools sufficiently, stars will form, which will then produce supernovae, which in turn will heat the ambient medium \citep{BinneyCoolevolution}. While this effect is certainly important, the amount of energy released can only be sufficient to offset cooling in low-luminosity elliptical galaxies, but does not even come close to the radiative losses in groups or cluster, or even high-luminosity ellipticals. Even in normal elliptical galaxies, supernova feedback models have been shown to need fine-tuning to produce gas halos that resemble X-ray observations \citep{MathewsReview}. 

Instead, the central active galactic nucleus (AGN) has emerged as the most popular heat source in the last few years, though this still remains an active matter of debate in the community. \citet{McNamaraAGNReview} give a recent review of AGN heating in clusters. {\it Chandra} and {\it XMM-Newton} observations \citep[e.g.][]{McNamaraHydraA} have revealed the existence of giant cavities in clusters, which are often associated with radio emission from ongoing jet activity of the central AGN, actively inflating these ``bubbles''. This direct link has also been observed on the smaller group \citep[e.g.][]{MoritaHCG62} and galaxy scales \citep[e.g.][]{Fin01,DiehlGallery,DiehlAGN,DiehlEntropy}.
AGN outbursts have been shown to have potentially more than sufficient energy to offset cooling in a large fraction of clusters \citep{BirzanCavities}, and even in normal elliptical galaxies and groups \citep{BestRadioloudAGN, BestAGNcooling}.

However, while the AGN has {\it potentially} sufficient energy to offset cooling, the detailed dynamics of how the released energy can actually be transferred into thermal energy of the gas is poorly understood. 
A lot of theoretical work has been undertaken in the purely hydrodynamical regime \citep[e.g.][]{BrueggenJets} to understand jet dynamics in inflating these cavities, but with limited success so far \citep[see e.g.][]{VernaleoJetProblems}. The most realistic simulations to date have been conducted by \citet{HeinzJets}, who employ the dentist's drill model combined with a dynamical cluster initial setup to avoid some of the issues other more idealized simulations have. But it remains to be seen how successful this model is on the less dynamic group and galaxy scale, where AGN heating has recently been found to be important as well \citep[e.g.][]{DiehlGallery, DiehlAGN, BestRadioloudAGN, BestAGNcooling, FabianAccretionJet, ShurkinNGC1399_4649} 

A problem of purely hydrodynamic jet simulations is that they ubiquitously produce a shock at the termination point of the jet, which results in hot, shocked gas surrounding the jet lobe. This is contrary to the vast majority of X-ray observations, though shocks around radio lobes have recently been reported in NGC~3801 \citep{CrostonShocks} and Centaurus~A \citep{KraftCentaurusA, KraftCentaurusA2}. Another major issue is that the created cavities often do not morphologically resemble X-ray observations, and get shredded apart quickly. Rayleigh-Taylor and Richtmyer-Meshkov instabilities are expected to grow on the top of the bubble and Kelvin-Helmholtz instabilities develop on the shear flow at the bubble sides, as it rises. Nevertheless, bubbles with rise times on the order of several times the predicted instability timescales have been observed, calling this scenario into question. Something has to suppress the growth of these instabilities. In general, tangential magnetic fields \citep{DeYoungMHDBubbles}, viscosity \citep{ReynoldsViscousBubble}, or continuous inflation of the bubble \citep{SokerRT} have been proposed to suppress the growth of instabilities.

Alternatively, \citet{LiMHD1} and \citet{LiMHD2, LiMHD3} have proposed a magnetically dominated jet model instead. Their 3D magneto-hydrodynamic (MHD) simulations advocate a current-carrying jet that evolves purely by injecting magnetic flux and energy into a small volume surrounding the central black hole. The injected magnetic fields are not force-free and interact with the ambient medium to form a self-collimated jet. As the background pressure drops, the jet expands into a wide jet lobe, excavating a bubble in the ambient medium. Mock X-ray observations of these systems show a strong resemblance to observations of cavities in clusters \citep{DiehlAlaska}. In particular, these cavities generally expand subsonically and appear cooler than the surrounding medium, similar to observations.

This paper addresses the question of whether one can use the bubbles' sizes  as they rise in the intracluster medium as a function of their distance to the cluster center as an indicator for which model may indeed be correct. 

The paper is organized as follows: In section \ref{s.data} we describe the methodology on how we identify the cavities, and extract their main properties. \S\ref{s.models} then presents the predictions of the various models that we are considering, and focus on their predictions for bubble sizes as a function of radius. In section \ref{s.observations} we show how this relates to observations. We then discuss various incompleteness issues associated with the data in section \ref{s.discussion}, before we try to reproduce the properties of our data set with Monte-Carlo simulations in \S\ref{s.montecarlo} and conclude (\S\ref{s.conclusions}).

\section{X-ray Data}\label{s.data}

\subsection{Cavity Sample}

Our cavity sample is mostly congruent with that of \citet{RaffertyBHgrowth}. The only differences are that we exclude  Abell~1068, which does not contain any cavities, and add the multiple cavities in Hydra~A from \citet{WiseHydraA}. Overall, the sample comprises a set of 64 individual cavities in 32 different hosts. Most of the hosts are clusters (30/32), with one group (HCG~62) and one individual galaxy (M~84). The cluster sample incorporates very poor clusters (e.g. Abell~262) to very rich clusters (e.g. Abell~1795). For clarity and simplicity, we will refer to the hosts of our cavity sample in general as ``clusters'', despite the two exceptions noted above. All of conclusions are not changed even if we exclude the one galaxy and the one group.

\subsection{Data Analysis}
The X-ray data, used to derive the cavity properties, are taken from the Chandra Data Archive and reprocessed in CIAO 3.3, as described in \citet{RaffertyBHgrowth}. In summary, we start from level 1 event files and apply a uniform calibration across the sample using CALDB 3.2.0. All systems have been observed with the ACIS detector in non-grating mode. During reprocessing, corrections to account for the time-dependent charge transfer inefficiency (CTI) and gain problems are applied. Additionally, the data are filtered for bad grades and hot pixels, and cosmic-ray events are identified and flagged. Lastly, exposure maps are made using weights appropriate for a single-temperature plasma.

Blank-sky background files\footnote{See http://cxc.harvard.edu/contrib/maxim/bg/} are used to estimate the background, as the cluster emission often fills the entire field of view of the CCD. Periods of high background are excluded from the source observation by filtering the data for periods during which the count rate exceeds 20\% the quiescent background count rate, after exclusion of all sources. The blank-sky background rate is normalized by the count rate in the 10-12 keV energy range and reprojected to match the aspect solution of the source data. After reprojection, the background spectra are extracted from the same regions of the CCD as the source spectra.

\subsection{Extraction of Cluster Properties}
For this work, we additionally extract X-ray surface brightness profiles using elliptical annuli centered on the X-ray core. The geometry of the annuli is fixed to that of the best-fit annulus at an intermediate radius, generally beyond the location of the cavities or other structure where the cluster emission is fairly smooth. The profiles are extracted between 0.1-12 keV and corrected for exposure with the exposure maps. 

The surface brightness profiles are then logarithmically binned to avoid that the model fits are driven mainly by noisy data at large radii. We fit a $\beta$-model profile with an additional background constant to each surface brightness profile by minimizing the $\chi^2$-value with the publicly available IDL\footnote{\texttt{http://www.ittvis.com/idl/}} fitting package MPFIT\footnote{\texttt{http://cow.physics.wisc.edu/$\sim$craigm/idl/fitting.html}}. For more information on the $\beta$-model, please refer to \citet{Sarazin_beta} or section \ref{s.betamodel}. Host properties are summarized in Table \ref{t.clusterprop}.

\subsection{Extraction of Bubble Properties}
We identify and measure 58 bubbles in the X-ray data \citep[in addition to the the six cavities in Hydra A identified and measured by][]{WiseHydraA}. Each bubble is confirmed as likely to be associated with the central AGN by comparing its location to that of the AGN's radio lobes \citep[see][]{birzan2007}. We note that most of the bubbles have been previously identified in the literature; see \citet{RaffertyBHgrowth} for references. In most cases, there is a direct anti-correlation between the X-ray and radio emission, such that the radio lobes fill cavities in the X-ray emission.

We measure the bubble properties following the procedure used in \citet {BirzanCavities,RaffertyBHgrowth}. Briefly, we estimate the size and position of each cavity by overlaying ellipses onto the exposure-corrected, unsmoothed images. The ellipses are positioned such that they encompass the cavity, but not any bright emission associated with the cavity rim, if present.

To obtain the density and temperature at the bubble's projected location, we perform a deprojection of the cluster emission using the PROJCT model in XSpec 11.3.1. Spectra of at least 3000 counts in the 0.5-7 keV band are extracted in concentric annuli, along with the associated weighted responses, using standard CIAO tools. The spectra are then fit with models of a single-temperature plasma (MEKAL) with foreground absorption (WABS) between the energies of 0.5-7 keV.  The foreground column density is held fixed to the Galactic value of \citet{DickeyHI}, and the redshift of the MEKAL component is fixed to the values given in \citet{RaffertyBHgrowth}. The temperature and abundance \citep[relative to the solar values of][]{AndersAbundances} of the MEKAL component are allowed to vary freely in each shell, and the density is calculated from the normalization of the MEKAL component assuming $n_e=1.2n_{\rm H}$.  Finally, the pressure is calculated as $p=nkT,$ where we assume a fully ionized ideal gas. The ambient pressure and temperature at the bubble's location are derived by interpolating the pressure and temperature profiles at the projected radius of the bubble's center. Bubble properties are summarized in Table \ref{t.bubbleprop}.

\begin{deluxetable*}{llccccccccc}
%\rotate
\tablewidth{0pt}
\tablecaption{Cluster X-ray Properties.\label{t.clusterprop}}
\tablehead{
\colhead{} & \colhead{Cluster} &  \colhead{$\tau_{\rm exp}$} & \colhead{$d$} & \colhead{${\rm k}T_0$} & \colhead{$n_0$} & \colhead{$p_0$} & \colhead{$r_c$} & \colhead{$\beta$} & \colhead{$M_{\rm BH}$} & \colhead{$L_{\rm NVSS}$}\\
 &  & [ks] & [Mpc] & [$\kev$] & [$10^{-2}$ & [$10^{-11}$& [$\kpc$] & & $[M_\sun]$ & [$\erg\, \sec^{-1}$]\\
 &  & & & & ${\rm cm}^{-3}$] & ${\rm erg} \, {\rm cm}^{-3}$]& & & &
}
\startdata
 1 & A85 & $ 37.5$ & $ 245.40$ & $ 2.06$ & $ 2.90$ & $ 1.91$ & $ 19.05$ & $0.41$ & $ 1.14\times 10^{9} $ & $3.38 \times 10^{29} $ \\
 2 & A133 & $ 30.9$ & $ 268.70$ & $ 1.77$ & $ 3.13$ & $ 1.78$ & $ 14.31$ & $0.42$ & \nodata & $2.68 \times 10^{29}$ \\
 3 & A262 & $ 26.7$ & $ 65.00$ & $ 0.86$ & $ 4.83$ & $ 1.33$ & $ 2.63$ & $0.36$ & $ 3.58\times 10^{8} $ & $3.32 \times 10^{29}$  \\
 4 & Perseus & $ 9.0$ & $ 78.20$ & $ 4.40$ & $ 3.13$ & $ 4.41$ & $ 37.58$ & $0.66$ & $ 3.15\times 10^{8} $  & $1.67 \times 10^{32}$\\
 5 & 2A0335 & $ 16.1$ & $ 153.90$ & $ 1.39$ & $ 3.68$ & $ 1.64$ & $ 19.78$ & $0.52$ & \nodata & \nodata  \\
 6 & A478 & $ 41.0$ & $ 368.10$ & $ 2.68$ & $ 2.87$ & $ 2.47$ & $ 30.75$ & $0.46$ & \nodata & \nodata\\
 7 & MS0735 & $ 39.9$ & $1068.30$ & $ 3.18$ & $ 2.76$ & $ 2.81$ & $ 20.86$ & $0.43$ & \nodata  & $ 2.88 \times 10^{31}$ \\
 8 & PKS0745 & $ 17.4$ & $ 475.10$ & $ 2.62$ & $ 3.44$ & $ 2.89$ & $ 34.32$ & $0.51$ & \nodata  & $ 6.40 \times 10^{32}$ \\
 9 & 4C55 & $ 66.5$ & $1214.50$ & $ 3.29$ & $ 2.59$ & $ 2.73$ & $ 33.21$ & $0.53$ & \nodata  & $1.46 \times 10^{34}$ \\
 10 & RBS797 & $ 8.7$ & $1856.90$ & $ 4.56$ & $ 2.92$ & $ 4.26$ & $ 57.38$ & $0.64$ & \nodata & $8.94 \times 10^{31}$  \\
 11 & Zw2701 & $ 22.2$ & $1057.20$ & $ 3.29$ & $ 1.58$ & $ 1.66$ & $ 49.72$ & $0.57$ & \nodata & $1.94 \times 10^{31}$  \\
 12 & Zw3146 & $ 41.5$ & $1493.20$ & $ 3.09$ & $ 3.01$ & $ 2.98$ & $ 46.69$ & $0.57$ & \nodata  & $1.89 \times 10^{31}$ \\
 13 & M84 & $ 25.6$ & $ 15.00$ & $ 0.57$ & $ 5.25$ & $ 0.96$ & $ 1.58$ & $0.75$ & $ 6.70\times 10^{8} $ & $1.62 \times 10^{30}$  \\
 14 & M87 & $ 15.1$ & $ 18.00$ & $ 0.94$ & $10.27$ & $ 3.08$ & $ 1.73$ & $0.36$ & $ 1.15\times 10^{9} $ & $5.36 \times 10^{31}$  \\
 15 & Centaurus & $ 48.9$ & $ 47.50$ & $ 0.77$ & $ 8.48$ & $ 2.08$ & $ 2.35$ & $0.37$ & $ 3.70\times 10^{8} $ & \nodata \\
 16 & HCG62 & $ 47.5$ & $ 60.60$ & $ 0.67$ & $ 4.10$ & $ 0.88$ & $ 3.59$ & $0.51$ & \nodata & \nodata \\
 17 & A1795 & $ 14.2$ & $ 282.70$ & $ 2.74$ & $ 2.58$ & $ 2.26$ & $ 35.16$ & $0.48$ & $ 6.35\times 10^{8} $ & \nodata\\
 18 & A1835 & $ 10.3$ & $1277.40$ & $ 4.03$ & $ 4.69$ & $ 6.05$ & $ 30.99$ & $0.52$ & \nodata & $7.67 \times 10^{31}$ \\
 19 & PKS1404 & $ 6.6$ & $ 95.80$ & $ 1.35$ & $ 2.77$ & $ 1.20$ & $ 9.26$ & $0.44$ & $ 3.75\times 10^{8} $ & $7.08 \times 10^{30}$  \\
 20 & MACS1423 & $110.2$ & $3142.30$ & $ 3.74$ & $ 3.67$ & $ 4.40$ & $ 27.43$ & $0.55$ & \nodata  & $9.44 \times 10^{31}$ \\
 21 & A2029 & $ 77.3$ & $ 348.90$ & $ 2.88$ & $ 2.93$ & $ 2.70$ & $ 36.00$ & $0.46$ & $ 1.53\times 10^{9} $ & \nodata \\
 22 & A2052 & $ 36.1$ & $ 153.90$ & $ 0.71$ & $ 1.87$ & $ 0.43$ & $ 22.20$ & $0.50$ & $ 3.81\times 10^{8} $ & \nodata \\
 23 & MKW3S & $ 55.2$ & $ 199.30$ & $ 2.81$ & $ 1.48$ & $ 1.33$ & $ 30.21$ & $0.45$ & \nodata & $5.46 \times 10^{30}$  \\
 24 & A2199 & $ 15.9$ & $ 131.40$ & $ 2.20$ & $ 2.25$ & $ 1.58$ & $ 21.35$ & $0.41$ & $ 7.07\times 10^{8} $ & \nodata \\
 25 & HercA & $ 12.5$ & $ 734.00$ & $ 1.98$ & $ 0.49$ & $ 0.31$ & $ 156.08$ & $0.95$ & \nodata & $1.74 \times 10^{34}$  \\
 26 & 3C388 & $ 25.8$ & $ 421.20$ & $ 3.02$ & $ 0.94$ & $ 0.91$ & $ 24.60$ & $0.46$ & $ 1.52\times 10^{9} $ & $1.19 \times 10^{33}$  \\
 27 & 3C401 & $ 21.8$ & $ 985.50$ & $ 2.76$ & $ 1.02$ & $ 0.90$ & $ 24.34$ & $0.50$ & \nodata & $5.88 \times 10^{33}$  \\
 28 & CygnusA & $ 34.2$ & $ 250.00$ & $ 5.25$ & $ 2.73$ & $ 4.59$ & $ 26.30$ & $0.57$ & \nodata & $1.19 \times 10^{35}$  \\
 29 & Sersic159 & $ 9.7$ & $ 259.30$ & $ 1.83$ & $ 2.53$ & $ 1.48$ & $ 31.98$ & $0.57$ & \nodata & \nodata \\
 30 & A2597 & $ 11.3$ & $ 387.30$ & $ 1.57$ & $ 3.27$ & $ 1.64$ & $ 31.76$ & $0.59$ & $ 2.05\times 10^{8} $ & $3.36 \times 10^{32}$  \\
 31 & A4059 & $ 92.1$ & $ 213.10$ & $ 2.08$ & $ 0.81$ & $ 0.54$ & $ 52.31$ & $0.51$ & $ 6.52\times 10^{8} $ & \nodata\\
 32 & HydraA & $ 98.8$ & $ 231.50$ & $ 2.59$ & $ 3.13$ & $ 2.60$ & $ 19.95$ & $0.47$ & $ 9.15\times 10^{8} $ & $2.62 \times 10^{33}$  \\
\enddata
\tablecomments{Column 1 -- Total ACIS exposure time in ksec; Column 2 -- Distance to the cluster in Mpc; Column 3 -- Central deprojected gas temperature ($\kev$); Column 4 -- Central density ($10^{-2}{\rm cm}^{-3}$); Column 5 -- Central pressure ($10^{-11}{\rm dyne} \, {\rm cm}^{-2}$); Column 6+7 -- $\beta$-profile fitted parameters, core radius and $\beta$ parameter; Column 8 -- black hole masses, derived from velocity dispersion data from Hyperleda \citep{Hyperleda} using the relation given by \citet{TremaineBHsigma}; Column 9 -- Integrated $20\cm$ Radio luminosity within $1\arcmin$ taken from the NVSS catalog \citep{NVSS}. }
\end{deluxetable*}

\begin{deluxetable*}{llccccccccccccccc}
%\rotate
\tablewidth{0pt}
\tablecaption{Cavity X-ray Properties.\label{t.bubbleprop}}
\tablehead{
\colhead{Nr.} & \colhead{Host} & \colhead{$r$} & \colhead{$R_{\rm b}$} & \colhead{$pV$} & \colhead{$I_z$} & \colhead{$\eta^{-1} \dot{E_0}$} & \colhead{$t_{\rm rise}$} & \colhead{$t_{\rm KH,max}$} & \colhead{$t_{\rm RT, max}$} &  \colhead{$t_{\rm RM, max}$}  \\
\colhead{} & \colhead{} & \colhead{[$\kpc$]} & \colhead{[$\kpc$]} & \colhead{[$\erg$]} & \colhead{[$10^{18}\,{\rm A}$]} & \colhead{$\erg \sec^{-1}$} & \colhead{[Myr]} & \colhead{[Myr]} & \colhead{[Myr]} &  \colhead{[Myr]}  
}
\startdata
  1 & A85                     & $ 21.3$ & $ 7.5$  & $ 6.03\times 10^{57} $ & $ 6.26 $ & $ 8.4 \times 10^{42} $ & $  25.7 $ & $ 920.8$  & $ 5.7$  & $ 1.2$ \\
  2 & A133                    & $ 32.4$ & $ 29.1$ & $ 1.70\times 10^{59} $ & $ 16.9 $ & $ 1.7 \times 10^{44} $ & $  42.1 $ & $ 3850.9$ & $ 11.9$ & $ 4.9$ \\
  3 & A262                    & $ 8.7$  & $ 4.3$  & $ 3.40\times 10^{56} $ & $ 1.96 $ & $ 9.3 \times 10^{41} $ & $  16.2 $ & $ 815.7$  & $ 3.5$  & $ 1.0$ \\
  4 & A262                    & $ 8.1$  & $ 4.4$  & $ 3.95\times 10^{56} $ & $ 2.09 $ & $ 1.1 \times 10^{42} $ & $  15.1 $ & $ 838.1$  & $ 3.5$  & $ 1.1$ \\
  5 & Perseus                 & $ 9.4$  & $ 8.2$  & $ 2.77\times 10^{58} $ & $ 12.9 $ & $ 1.1 \times 10^{44} $ & $   7.8 $ & $ 686.0$  & $ 6.6$  & $ 0.9$ \\
  6 & Perseus                 & $ 6.5$  & $ 6.2$  & $ 1.26\times 10^{58} $ & $ 9.94 $ & $ 7.5 \times 10^{43} $ & $   5.4 $ & $ 522.5$  & $ 6.8$  & $ 0.7$ \\
  7 & Perseus                 & $ 28.3$ & $ 11.1$ & $ 4.83\times 10^{58} $ & $ 14.5 $ & $ 7.3 \times 10^{43} $ & $  23.4 $ & $ 937.6$  & $ 5.4$  & $ 1.2$ \\
  8 & Perseus                 & $ 38.6$ & $ 14.6$ & $ 8.29\times 10^{58} $ & $ 16.6 $ & $ 9.8 \times 10^{43} $ & $  31.9 $ & $ 1226.9$ & $ 6.0$  & $ 1.6$ \\
  9 & 2A0335                  & $ 23.1$ & $ 7.8$  & $ 4.82\times 10^{57} $ & $ 5.49 $ & $ 5.3 \times 10^{42} $ & $  33.9 $ & $ 1164.6$ & $ 6.4$  & $ 1.5$ \\
 10 & 2A0335                  & $ 27.5$ & $ 3.5$  & $ 3.81\times 10^{56} $ & $ 2.29 $ & $ 3.7 \times 10^{41} $ & $  40.4 $ & $ 529.2$  & $ 4.4$  & $ 0.7$ \\
 11 & A478                    & $ 9.0$  & $ 4.3$  & $ 2.32\times 10^{57} $ & $ 5.11 $ & $ 7.8 \times 10^{42} $ & $   9.5 $ & $ 466.5$  & $ 6.2$  & $ 0.6$ \\
 12 & A478                    & $ 9.0$  & $ 4.4$  & $ 2.38\times 10^{57} $ & $ 5.15 $ & $ 8.0 \times 10^{42} $ & $   9.5 $ & $ 470.7$  & $ 6.2$  & $ 0.6$ \\
 13 & MS0735                  & $156.5$ & $ 98.0$ & $ 2.43\times 10^{60} $ & $ 34.7 $ & $ 9.7 \times 10^{44} $ & $ 151.9 $ & $ 9707.4$ & $ 33.1$ & $ 12.4$ \\
 14 & MS0735                  & $183.4$ & $108.0$ & $ 2.66\times 10^{60} $ & $ 34.6 $ & $ 9.5 \times 10^{44} $ & $ 178.0 $ & $10697.4$ & $ 37.5$ & $ 13.7$ \\
 15 & PKS0745                 & $ 30.6$ & $ 20.9$ & $ 2.09\times 10^{59} $ & $ 22.0 $ & $ 2.3 \times 10^{44} $ & $  32.8 $ & $ 2286.1$ & $ 10.2$ & $ 2.9$ \\
 16 & 4C55                    & $ 16.1$ & $ 8.8$  & $ 1.92\times 10^{58} $ & $ 10.3 $ & $ 4.1 \times 10^{43} $ & $  15.4 $ & $ 856.0$  & $ 6.4$  & $ 1.1$ \\
 17 & 4C55                    & $ 21.6$ & $ 11.1$ & $ 3.46\times 10^{58} $ & $ 12.3 $ & $ 5.7 \times 10^{43} $ & $  20.6 $ & $ 1080.7$ & $ 6.6$  & $ 1.4$ \\
 18 & RBS797                  & $ 23.8$ & $ 10.7$ & $ 5.47\times 10^{58} $ & $ 15.8 $ & $ 9.3 \times 10^{43} $ & $  19.3 $ & $ 882.4$  & $ 7.5$  & $ 1.1$ \\
 19 & RBS797                  & $ 19.5$ & $ 9.7$  & $ 4.31\times 10^{58} $ & $ 14.7 $ & $ 8.9 \times 10^{43} $ & $  15.8 $ & $ 802.0$  & $ 7.7$  & $ 1.0$ \\
 20 & Zw2701                  & $ 54.1$ & $ 43.5$ & $ 8.62\times 10^{59} $ & $ 31.0 $ & $ 6.3 \times 10^{44} $ & $  51.6 $ & $ 4235.4$ & $ 14.8$ & $ 5.4$ \\
 21 & Zw2701                  & $ 48.6$ & $ 36.7$ & $ 5.67\times 10^{59} $ & $ 27.4 $ & $ 4.5 \times 10^{44} $ & $  46.4 $ & $ 3567.7$ & $ 13.6$ & $ 4.6$ \\
 22 & Zw3146                  & $ 39.8$ & $ 33.2$ & $ 8.42\times 10^{59} $ & $ 35.1 $ & $ 7.7 \times 10^{44} $ & $  39.2 $ & $ 3329.8$ & $ 13.1$ & $ 4.3$ \\
 23 & Zw3146                  & $ 59.1$ & $ 33.0$ & $ 5.88\times 10^{59} $ & $ 29.4 $ & $ 3.9 \times 10^{44} $ & $  58.2 $ & $ 3317.7$ & $ 13.1$ & $ 4.2$ \\
 24 & M84                     & $ 2.3$  & $ 1.6$  & $ 1.34\times 10^{55} $ & $ 0.64 $ & $ 1.1 \times 10^{41} $ & $   5.3 $ & $ 374.8$  & $ 1.1$  & $ 0.5$ \\
 25 & M84                     & $ 2.5$  & $ 1.6$  & $ 1.15\times 10^{55} $ & $ 0.59 $ & $ 9.0 \times 10^{40} $ & $   5.7 $ & $ 371.9$  & $ 1.1$  & $ 0.5$ \\
 26 & M87                     & $ 2.8$  & $ 1.8$  & $ 1.09\times 10^{56} $ & $ 1.72 $ & $ 8.2 \times 10^{41} $ & $   5.0 $ & $ 327.4$  & $ 1.4$  & $ 0.4$ \\
 27 & M87                     & $ 2.2$  & $ 1.1$  & $ 3.26\times 10^{55} $ & $ 1.18 $ & $ 3.0 \times 10^{41} $ & $   3.9 $ & $ 206.5$  & $ 1.1$  & $ 0.3$ \\
 28 & Centaurus               & $ 6.0$  & $ 2.8$  & $ 1.86\times 10^{56} $ & $ 1.79 $ & $ 6.6 \times 10^{41} $ & $  11.9 $ & $ 567.7$  & $ 2.6$  & $ 0.7$ \\
 29 & Centaurus               & $ 3.5$  & $ 2.3$  & $ 1.62\times 10^{56} $ & $ 1.85 $ & $ 8.7 \times 10^{41} $ & $   6.9 $ & $ 463.5$  & $ 2.0$  & $ 0.6$ \\
 30 & HCG62                   & $ 8.4$  & $ 4.6$  & $ 2.57\times 10^{56} $ & $ 1.64 $ & $ 6.5 \times 10^{41} $ & $  17.8 $ & $ 1003.1$ & $ 3.6$  & $ 1.3$ \\
 31 & HCG62                   & $ 8.6$  & $ 4.0$  & $ 1.60\times 10^{56} $ & $ 1.40 $ & $ 4.0 \times 10^{41} $ & $  18.2 $ & $ 865.3$  & $ 3.4$  & $ 1.1$ \\
 32 & A1795                   & $ 18.5$ & $ 11.5$ & $ 3.60\times 10^{58} $ & $ 12.3 $ & $ 6.2 \times 10^{43} $ & $  19.4 $ & $ 1231.8$ & $ 8.5$  & $ 1.6$ \\
 33 & A1835                   & $ 23.3$ & $ 13.4$ & $ 1.27\times 10^{59} $ & $ 21.4 $ & $ 2.2 \times 10^{44} $ & $  20.1 $ & $ 1179.9$ & $ 6.3$  & $ 1.5$ \\
 34 & A1835                   & $ 16.6$ & $ 11.5$ & $ 9.28\times 10^{58} $ & $ 19.8 $ & $ 2.2 \times 10^{44} $ & $  14.3 $ & $ 1010.7$ & $ 6.3$  & $ 1.3$ \\
 35 & PKS1404                 & $ 4.6$  & $ 3.0$  & $ 3.49\times 10^{56} $ & $ 2.37 $ & $ 1.7 \times 10^{42} $ & $   6.8 $ & $ 459.0$  & $ 3.4$  & $ 0.6$ \\
 36 & PKS1404                 & $ 3.8$  & $ 2.9$  & $ 3.37\times 10^{56} $ & $ 2.36 $ & $ 1.9 \times 10^{42} $ & $   5.7 $ & $ 447.3$  & $ 3.5$  & $ 0.6$ \\
 37 & MACS1423                & $ 15.8$ & $ 9.4$  & $ 3.56\times 10^{58} $ & $ 13.6 $ & $ 8.5 \times 10^{43} $ & $  14.1 $ & $ 858.2$  & $ 5.3$  & $ 1.1$ \\
 38 & MACS1423                & $ 16.7$ & $ 9.4$  & $ 3.47\times 10^{58} $ & $ 13.4 $ & $ 7.9 \times 10^{43} $ & $  15.0 $ & $ 858.2$  & $ 5.2$  & $ 1.1$ \\
 39 & A2029                   & $ 32.1$ & $ 9.7$  & $ 2.01\times 10^{58} $ & $ 10.0 $ & $ 2.1 \times 10^{42} $ & $  32.7 $ & $ 1006.6$ & $ 7.1$  & $ 1.3$ \\
 40 & A2052                   & $ 11.2$ & $ 9.2$  & $ 3.39\times 10^{57} $ & $ 4.24 $ & $ 4.9 \times 10^{42} $ & $  23.0 $ & $ 1918.6$ & $ 11.7$ & $ 2.4$ \\
 41 & A2052                   & $ 6.7$  & $ 6.3$  & $ 1.26\times 10^{57} $ & $ 3.10 $ & $ 2.9 \times 10^{42} $ & $  13.8 $ & $ 1331.0$ & $ 11.7$ & $ 1.7$ \\
 42 & MKW3S                   & $ 58.7$ & $ 34.9$ & $ 2.42\times 10^{59} $ & $ 18.3 $ & $ 1.6 \times 10^{44} $ & $  60.6 $ & $ 3677.0$ & $ 14.0$ & $ 4.7$ \\
 43 & A2199                   & $ 18.9$ & $ 12.2$ & $ 2.49\times 10^{58} $ & $ 9.96 $ & $ 3.9 \times 10^{43} $ & $  22.1 $ & $ 1454.9$ & $ 7.5$  & $ 1.9$ \\
 44 & A2199                   & $ 21.2$ & $ 12.6$ & $ 2.56\times 10^{58} $ & $ 9.94 $ & $ 3.7 \times 10^{43} $ & $  24.7 $ & $ 1501.2$ & $ 7.5$  & $ 1.9$ \\
 45 & HercA                   & $ 60.3$ & $ 23.4$ & $ 3.98\times 10^{58} $ & $ 9.09 $ & $ 1.8 \times 10^{43} $ & $  74.3 $ & $ 2934.6$ & $ 23.4$ & $ 3.7$ \\
 46 & HercA                   & $ 57.6$ & $ 29.7$ & $ 8.30\times 10^{58} $ & $ 11.7 $ & $ 3.9 \times 10^{43} $ & $  70.9 $ & $ 3730.2$ & $ 26.9$ & $ 4.8$ \\
 47 & 3C388                   & $ 26.8$ & $ 15.2$ & $ 2.30\times 10^{58} $ & $ 8.57 $ & $ 3.1 \times 10^{43} $ & $  26.7 $ & $ 1545.2$ & $ 7.2$  & $ 2.0$ \\
 48 & 3C388                   & $ 21.2$ & $ 15.4$ & $ 2.81\times 10^{58} $ & $ 9.41 $ & $ 4.6 \times 10^{43} $ & $  21.1 $ & $ 1569.5$ & $ 7.3$  & $ 2.0$ \\
 49 & 3C401                   & $ 15.0$ & $ 12.1$ & $ 1.55\times 10^{58} $ & $ 7.88 $ & $ 3.3 \times 10^{43} $ & $  15.6 $ & $ 1285.4$ & $ 6.8$  & $ 1.6$ \\
 50 & 3C401                   & $ 15.3$ & $ 12.1$ & $ 1.53\times 10^{58} $ & $ 7.85 $ & $ 3.2 \times 10^{43} $ & $  15.9 $ & $ 1285.4$ & $ 6.7$  & $ 1.6$ \\
 51 & CygnusA                 & $ 43.0$ & $ 22.3$ & $ 2.07\times 10^{59} $ & $ 21.2 $ & $ 2.7 \times 10^{44} $ & $  32.5 $ & $ 1721.5$ & $ 6.5$  & $ 2.2$ \\
 52 & CygnusA                 & $ 44.7$ & $ 27.9$ & $ 3.85\times 10^{59} $ & $ 25.9 $ & $ 4.8 \times 10^{44} $ & $  33.8 $ & $ 2153.8$ & $ 7.3$  & $ 2.7$ \\
 53 & Sersic159               & $ 23.2$ & $ 16.8$ & $ 6.00\times 10^{58} $ & $ 13.2 $ & $ 7.0 \times 10^{43} $ & $  29.7 $ & $ 2190.3$ & $ 10.2$ & $ 2.8$ \\
 54 & Sersic159               & $ 26.4$ & $ 18.9$ & $ 7.88\times 10^{58} $ & $ 14.2 $ & $ 8.3 \times 10^{43} $ & $  33.8 $ & $ 2466.1$ & $ 10.6$ & $ 3.1$ \\
 55 & A2597                   & $ 23.1$ & $ 7.1$  & $ 4.96\times 10^{57} $ & $ 5.83 $ & $ 5.4 \times 10^{42} $ & $  31.9 $ & $ 1001.0$ & $ 7.0$  & $ 1.3$ \\
 56 & A2597                   & $ 22.6$ & $ 8.5$  & $ 8.66\times 10^{57} $ & $ 7.03 $ & $ 9.6 \times 10^{42} $ & $  31.2 $ & $ 1199.8$ & $ 7.7$  & $ 1.5$ \\
 57 & A4059                   & $ 22.7$ & $ 14.4$ & $ 1.73\times 10^{58} $ & $ 7.65 $ & $ 2.1 \times 10^{43} $ & $  27.3 $ & $ 1758.8$ & $ 13.6$ & $ 2.2$ \\
 58 & A4059                   & $ 19.3$ & $ 9.2$  & $ 4.71\times 10^{57} $ & $ 4.99 $ & $ 6.6 \times 10^{42} $ & $  23.2 $ & $ 1127.3$ & $ 11.5$ & $ 1.4$ \\
 59 & HydraA\tablenotemark{*} & $ 24.1$ & $ 15.4$ & $ 6.21\times 10^{58} $ & $ 14.0 $ & $ 8.9 \times 10^{43} $ & $  25.9 $ & $ 1689.3$ & $ 7.0$  & $ 2.2$ \\
 60 & HydraA\tablenotemark{*} & $ 24.7$ & $ 15.5$ & $ 6.22\times 10^{58} $ & $ 14.0 $ & $ 8.7 \times 10^{43} $ & $  26.6 $ & $ 1702.9$ & $ 7.0$  & $ 2.2$ \\
 61 & HydraA\tablenotemark{*} & $ 97.4$ & $ 37.2$ & $ 1.72\times 10^{59} $ & $ 15.0 $ & $ 9.2 \times 10^{43} $ & $ 104.7 $ & $ 4085.6$ & $ 17.2$ & $ 5.2$ \\
 62 & HydraA\tablenotemark{*} & $ 57.3$ & $ 23.8$ & $ 8.99\times 10^{58} $ & $ 13.6 $ & $ 6.8 \times 10^{43} $ & $  61.6 $ & $ 2608.5$ & $ 10.9$ & $ 3.3$ \\
 63 & HydraA\tablenotemark{*} & $217.9$ & $ 98.8$ & $ 1.06\times 10^{60} $ & $ 22.8 $ & $ 3.3 \times 10^{44} $ & $ 234.3 $ & $10841.0$ & $ 41.3$ & $ 13.8$ \\
 64 & HydraA\tablenotemark{*} & $100.7$ & $ 56.3$ & $ 5.67\times 10^{59} $ & $ 22.1 $ & $ 3.0 \times 10^{44} $ & $ 108.3 $ & $ 6170.8$ & $ 21.5$ & $ 7.9$ \\
\enddata
\tablenotetext{*}{Hydra~A data taken from \citet{WiseHydraA}}
\tablecomments{Column 1+2 == Cavity number and host name; Column 3 -- Projected distance of the bubble center to the centroid of the cluster X-ray emission; Column 4 -- Size of the bubble; Column 5 -- $pV$ work associated with inflating a spherical bubble of size $V=4/3\pi R_b^3$ at $r$; Column 6 -- Current for CDJ model (from equation \ref{e.current}); Column 7 -- Energy injection rate for the CIH model (from equation \ref{e.rb_inflated}); Column 8 -- Bubble age assuming a constant rise speed at the speed of sound; Column 9,10,11 -- Kelvin-Helmholtz, Rayleigh-Taylor and Richtmyer-Meshkov instability timescales for the largest possible mode ($R_b$).}
\end{deluxetable*}

%%%%%%%%%%%%%%%%%%%%%%%%%%%%%%%%%%%%%

\section{Theory: Predictions for Bubble Sizes as a Function of Radius}\label{s.models}

\subsection{Assumptions and Underlying Thermodynamic Structure}\label{s.betamodel}

To distinguish between different models of X-ray cavities, we need to compare definite predictions of the various models. One such well-defined prediction is the evolution of the bubble sizes as they rise in the intracluster medium. Since X-ray observations show no evidence for shocks around cavities, we assume throughout this paper that their inflation and expansion is subsonic and in approximate pressure equilibrium with the surrounding gas. In the following sections we will summarize the predictions of various models and show their predictions for the bubble expansion as a function of radius. 

To make these predictions, we have to assume a background pressure profile, which we choose to be the widely known and used $\beta$ model \citep[e.g.][]{Sarazin_beta}. This model adequately describes a wide range of cluster profiles over a rather wide range of radii, and its shape is uniquely defined by only two parameters: the core radius $r_c$ and the exponential parameter $\beta$, which determines the sharpness of the turnover beyond the core radius, as well as the asymptotic slope of the profile. The surface brightness profile can be described as
\begin{equation}
  S(R)={S_0} {\left[ 1+ (R/r_c)^2 \right]^{-3\beta-{1\over2}} }, 
 \end{equation}
where $R$ denotes the projected radius.\footnote{This corresponds to the profile assumed in \citet{LiMHD3} with $\kappa=3\beta/2$ as used in their MHD simulations.}
 
One of the most advantageous properties of the $\beta$ model is that one can analytically derive its deprojection into three dimensions for the case of perfect spherical symmetry, which has made this model the de-facto standard model used for clusters in X-ray astronomy. Assuming isothermality at a temperature $T$ throughout the cluster then allows us to determine the density and pressure profile as a function of the spherical radius $r$ \citep[e.g.][]{EttoriBetamodel} : 
\begin{equation}
  \rho(r)= \rho_0 \left[ 1+ (r/r_c)^2 \right]^{-{3\beta\over 2}} {\rm,\,and} 
\end{equation}
\begin{equation}
  p_{\rm gas}(r)=p_0 {\left[ 1+ (r/r_c)^2 \right]^{-{3\beta\over 2}} }.
\end{equation}
Here, the central pressure $p_0$ is connected to the central density $\rho_0$ and isothermal temperature $T$ and average molecular weight $\bar{m}$ via the equation of state of an ideal gas with an adiabatic index $\Gamma=5/3$: $p_0=\rho_0\,\bar{m}^{-1}\,kT$. Though this represents a simplified picture of a real cluster, it allows us to make analytical predictions which should at least qualitatively hold up even in more complex systems. Real clusters tend to exhibit cool cores, an outwardly rising temperature gradient, and sometimes also an additional surface brightness excess near the center. We did repeat the analysis that follows with the more complex double-$\beta$ and temperature models, which leaves our conclusions unchanged and results in no additional insight into the problem. Thus, we decided to discuss only the isothermal single-$\beta$ model for simplicity. About 10 clusters in our sample show signs of a second $\beta$ model in the very center. However, we made sure that the single-$\beta$ model fit adequately represented the pressure profile around the location of the cavity, which is all our discussion here requires. 

In the following, we will present predictions of various models with regard to the size evolution of cavities as they rise in this ambient pressure profile. We assume that they are always in approximate pressure equilibrium with the surroundings. This assumption is justified by the general absence of strong shocks surrounding the bubbles, though there is now evidence for weak shocks with Mach numbers ranging from $1.2$--$1.7$ in a number of systems \citep[see][for a recent review]{McNamaraAGNReview}. These weak shocks are usually not directly associated with the cavity rims however, but may rather indicate an earlier faster expansion phase of the bubble. Only two objects are known with evidence for shocks in direct contact with the radio lobes: Centaurus~A \citep{KraftCentaurusA,KraftCentaurusA2} and NGC~3801 \citep{CrostonShocks}. In fact, the majority of rims around bubbles seem to be cooler than the ambient medium, suggesting a subsonic inflation and rise of the bubbles, which provides a solid basis for our assumption. 

\subsection{Adiabatic Expansion of Purely Hydrodynamic Bubbles (AD43/AD53)}

The standard assumption in most hydrodynamic simulations of cluster-scale jets is that the jet is kinetically dominated and inflates a bubble at the location of the jet lobes. We assume that this bubble is filled with ideal gas with an adiabatic index $\Gamma$. For simplicity, we will assume that the bubble is inflated at the center for this model with an internal pressure $p_{b,0}$ and Volume $V_{b,0}=4/3\pi\, R_{b,0}^3$. How the total outburst energy $E_\Gamma$ is related to the mechanical work associated with inflating this bubble, is a matter of debate and can be parameterized by an efficiency parameter $\eta$ via $E_\Gamma=\eta\,p_{b,0}V_{b,0}$. Realistic 3-dimensional hydrodynamic simulations \citep[e.g.][]{HeinzJets} are needed to determine the value of $\eta$ and to find out whether it is a function of environmental parameters, cluster dynamics, black hole mass, or similar quantities. The initial bubble radius $R_{b,0}$ at the cluster center should be a direct function of the outburst energy and this efficiency parameter, such that
\begin{equation} \label{e.r0_gamma}
R_{b,0}=\left({3E_\Gamma \over 4\pi \eta p_0}\right)^{1/3}. 
\end{equation}

Our analysis is insensitive to the details of this mechanism. For now, let us simply assume that we have a purely hydrodynamic, inflated bubble which adiabatically expands while rising in the intracluster medium. This bubble should expand according to the well-known adiabatic law $p_b V_b^\Gamma=p_{b,0}V_{b,0}^{\Gamma}$. Combining this relation with the assumed thermodynamic structure of the cluster gas predicts the evolution of the bubble radius $R_{b,\Gamma}$ as a function of radius, assuming the bubble stays intact and is constantly in pressure equilibrium with the ambient medium:
\begin{equation}\label{e.rb_gamma}
  R_{b,\Gamma}=R_{b,0} \left[ 1+ (r/r_c)^2 \right]^{{\beta\over2\Gamma}}. 
\end{equation}\\
Thus, at large radii, the cavities should grow asymptotically as  $R_{b,\Gamma}\propto (r/r_c)^{-{\beta/\Gamma}}$. For typical values of $\beta=0.5$ for a cluster and a relativistic gas with $\Gamma=4/3$, we get $R_{b,\Gamma}\propto (r/r_c)^{3/8}$; for $\Gamma=5/3$ we get $R_{b,\Gamma}\propto (r/r_c)^{3/10}$. From here on, we will refer to this model as the adiabatic expansion model, or short AD43 and AD53 for $\Gamma=4/3$ and $5/3$, respectively. Both correlations are actually surprisingly shallow, especially as one should rather consider these bubble size predictions as upper limits, since they do not take the effects of hydrodynamic instabilities into account. In fact, we will show later in section \ref{s.instabilities} that these instabilities should have shredded most hydrodynamic bubbles apart. Hydrodynamic simulations including viscosity have recently been found to stabilize the bubbles and suppress the development of instabilities \citep{ReynoldsViscousBubble}. 

Another effect not included in these predictions is the radiative loss of pressure support. If the highly relativistic particles inside the cavities provide pressure support and radiate a significant fraction of their energy away, the predicted size evolution of the cavities would be even shallower. Typical radiative lifetimes for the relativistic electrons are around $10^6-10^7\yr$, on par with bubble ages. However, it has been shown \citep{DunnFillingFactor, birzan2007} that the relativistic electrons are insufficient to supply the pressure needed to keep the bubbles in pressure equilibrium. The missing pressure component must be either supplied by ``extra'' particles, such as protons or entrained thermal gas, or magnetic fields. 

\subsection{Continuously Inflated Hydrodynamic Bubbles (CIH)}

The just described AD43 and AD53 models both assume the existence of an a priori preformed bubble at the center which then buoyantly rises. In this extreme case, the outburst energy is deposited instantaneously by the AGN. In this section, we now consider the other extreme with a continuous energy injection over time. For simplicity, we assume this process to be constant (i.e. $E(t)=\dot{E_0}\,t$). Assuming the AGN injects purely hydrodynamic material, we can rewrite this as a change in initial volume $V_0$ with efficiency $\eta$ over time, rather than injected energy: $\dot{E_0}t = \eta p_0 V_0(t) = 4/3 \eta p_0 \pi R_b^3(t)$. Replacing $R_{b,0}$ in equation \ref{e.rb_gamma} with its new time dependent form, and assuming a continuous rise speed (e.g. at the sound speed, $c_s=r/t$), yields the prediction for the radial evolution of the bubble size for the continuously inflated hydrodynamic (CIH) bubble model: 
\begin{equation}\label{e.rb_inflated}
  R_{b, {\rm CIH}}=\left({{3\dot{E_0}r}\over{4\pi \eta p_0 c_s}}\right)^{1/3} \left[ 1+ (r/r_c)^2 \right]^{{\beta\over2\Gamma}}. 
\end{equation}\\
Asymptotically at large radii, this model predicts that the bubble size scales as $R_{b}\propto r^{17/24}$ for a $\Gamma=4/3$ and a $\beta$ parameter of $0.5$.

Note that this model implicitly relies on several basic assumptions. First, we assume in equation (\ref{e.rb_inflated}) that all of the new material that is injected by the AGN is directly incorporated into the cavity at radius $r$, and instantly adiabatically expanded to match the outside gas pressure. In a realistic model, this would require the jet ``nozzle'' to always be directly attached to the rising bubble, while keeping the bubble intact. Realistic 3D hydro simulations are needed to address the question whether this is actually feasible, without producing a hot spot for the majority of cavities. Current state-of-the-art simulations tend to suggest otherwise \citep[e.g.][]{VernaleoJetProblems, BrueggenJets}. Another problem of this model is the length of the injection time, which in our toy model would then be equal to the bubble ages $t_{\rm rise}$ given in Table \ref{t.bubbleprop}. These ages are are about an order of magnitude longer than what is expected for typical AGN life times and duty cycles \citep[e.g.][]{AdelbergerDutyCycle}. 

Thus, this CIH model should rather be considered the other ``extreme'' and realistic paths in the $r-R_b$ diagram should fall in between the CIH and AD models. In fact, one could realize arbitrary injection functions $\dot{E_0}(t)$, and accordingly replace the injection term in equation \ref{e.rb_inflated}. 

\subsection{Magnetically Dominated Bubbles with Flux-Frozen Magnetic Loops (FML)}

In the following, we show the prediction for the size evolution of a magnetically dominated bubble for the case of a flux-frozen magnetic dipole field, refering to this model as the FML model. We follow closely the discussion by \citet{ThompsonMagDipole}, who outline this effect for magnetic fields trapped during a core collapse of a star leading to a magnetized neutron star. The basic idea is that the bubble is threaded by magnetic fields, which are dragged along when the bubble expands during its rise in the cluster atmosphere. 

Let us consider a magnetic field configuration, in which the bubble surface is filled densely with magnetic flux loops, analogous to the surface of a star. We further assume that these loops have a characteristic length scale $l_{\rm loop}$, so that we can approximate their dipole moment as $\mu_{\rm loop} \approx 3\Phi_{\rm loop} l_{\rm loop}/8\pi$, where $\phi_{\rm loop}$ is the average magnetic flux of each loop. For a random orientation of the loops, the net dipole moment of the whole bubble can then be written as $\mu_{b}\approx N_{\rm loop}^{1\over 2}\mu_{\rm loop} \approx {3\over 4}\pi^{-{1\over 2}} \Phi R_{b,\mu}$. Thus, we can approximate the magnetic dipole field as $B_{\rm dipole}= {{3\over 2} \pi^{-{1\over 2}} \, \Phi_{\rm loop} R_{b,\mu}^{-2}}$, as given by equation (39) of \citet{ThompsonMagDipole}. Assuming that the magnetic field dominates the pressure of the bubble, we can then derive the size of a bubble as a function of radius by balancing the internal magnetic pressure by the external gas pressure. Thus, we can express the bubble radius as 
\begin{equation}\label{e.rb_dipole}
  R_{b,\mu}= R_{b,0} \left[ 1+ (r/r_c)^2 \right]^{{3\beta\over 8}}. \\
\end{equation}
with the initial bubble size projected to the cluster center given by
\begin{equation}\label{e.r0_dipole}
R_{b,0}=\left( {{8\pi\,\mu_b^2}\over{p_0} } \right)^{1 \over 6} .
\end{equation}

Note that the exponent in the FMD model is identical to that of an adiabatic bubble with an adiabatic index of $\Gamma=4/3$ (model AD43). Thus, one cannot use the size evolution of bubbles to distinguish between these two models. Neither can we use this information to constrain the fraction of pressure support that is provided from randomly oriented magnetic flux loops.

\subsection{Magnetically Dominated Bubbles from Current-Dominated Jets (CDJ)}

Three-dimensional magneto-hydrodynamic simulations of current-dominated jets (CDJ) have recently been modeled for the first time \citep{LiMHD1, LiMHD2,LiMHD3}. The jet in these simulations is launched purely by injecting non-force-free poloidal and toroidal magnetic fields. These fields then launch a magnetic tower, self-collimate the jet and subsonically expand a lobe structure close to the point where the background gas pressure starts to quickly drop beyond the core radius. 

This jet system is a connected, current-carrying system: the current $I_z$ first travels along the inner jet axis to the lobes, and then returns on the outside of the lobe and in a sheath around the jet axis. This results in a magnetic field structure that resembles a tightly wound helix inside the jet, with another helical field wrapped around, but of opposite sign and less tightly wound. The magnitude of the $\phi$-component of the magnetic field within the lobe is essentially given by applying Amp\`ere's law\footnote{We use cgs units throughout this paper.}: $B_\phi=2c^{-1} I_z\,R^{-1}$. Since the bubbles are magnetically dominated, this also sets the size of the bubble which is in quasi-pressure equilibrium with the ambient gas pressure:
\begin{equation}\label{e.rb_current}
  R_{b,{\rm I_z}}= R_{b,0} \left[ 1+ (r/r_c)^2 \right]^{{3\beta\over 4}} 
\end{equation}

The initial size $R_{b,0}$ of the bubble projected back to the cluster center is then a rather simple function of the current $I_z$ flowing through the system: 
\begin{equation}\label{e.rbinitmag}\label{e.r0_current}
R_{b,0}={1 \over \sqrt{2\pi} \, c}  \, p_0^{-{1 \over 2}} \, I_z 
\end{equation}
To get a better understanding of the magnitude of the currents that are expected to be involved, we consider a central black hole with a mass around $2\times 10^8 M_\sun$, with a Schwarzschild radius $r_{\rm BH}=2GM_{\rm BH}\,c^{-2}$ of about $5.9\times 10^{13} \cm$. Let us further assume an equipartition magnetic field of $10^4 \,{\rm Gauss}$ in the accretion disk around the black hole, that is being threaded through the black hole. The current can then be estimated as $I_z=\pi r^2\, j$ with the current density $j=(c/4\pi)\, \nabla\times {\bf B}$.  Thus, the current\footnote{Note that we used the more commonly used non-cgs unit Amp\`ere instead of ${\rm esu}\s^{-1}$ ($1{\rm A}=2.998\times 10^{9}{\rm esu}\, \s^{-1}$).} can be approximated as $I_z=(c/4)\, B\,r$, yielding approximately $1.5\times 10^{18}\,A$ for a $2\times 10^8 M_\sun$ black hole. For more details, please refer to \citet{LiMHD1}.

Thus, we can now express equation (\ref{e.rbinitmag}) as a function of typical values found in galaxy clusters. For a central electron number density of $n_0=10^{-2}\cm^{-3}$ and an average temperature of ${\rm k}T_0=5\kev$, we get an initial bubble size of $R_{b,0}=1.021\kpc$ for a driving current of $I_z=10^{18}A$:
\begin{equation}\label{e.current}
R_{b,0}	=  1.021\kpc  \left( {n_0 \over 10^{-2}\cm^{-3}} \right)^{-{1 \over 2} }  \, \left( {{\rm k}T \over 5 \kev}\right)^{-{1 \over 2} } \, \left( {I_z \over 10^{18} \,{\rm A}} \right).
\end{equation}

%I = pi*r^2 * j
%j = (c/4pi)*curl(B) = (c/4pi)*B/r
%so, I = (c/4)*B*r

\subsection{Summary of Models}

All of the presented models predict that the bubble size evolves in a rather simple analytic form: 
\begin{equation}
R_b=R_{b,0}\, \left[ 1+ (r/r_c)^2 \right]^{\alpha}. 
\end{equation}
Thus, the behavior of the models is very similar, and they can be characterized by only two parameters: they start with an initial bubble size $R_{b,0}$ projected back to the cluster center, and expand according to the exponent $\alpha$. Table~\ref{t.models} summarizes these parameters for all models. Note that $R_{b,0}$ is related to the energetics of the jet or outburst and the central pressure, while $\alpha$ distinguishes the nature of the bubbles by the steepness of the evolution during the rise. Table \ref{t.models} is organized in increasing order of $\alpha$, i.e. the purely hydrodynamic model with $\Gamma=5/3$ (AD53) is the shallowest, while the magnetically dominated bubbles in the current-dominated jet (CDJ) model expand the fastest.

\begin{deluxetable*}{llcc}
\tablewidth{0pt}
\tablecaption{Comparison of bubble size evolution models. Please refer to section \ref{s.models} for full information on the parameters. \label{t.models}}
\tablehead{
\colhead{Model} & \colhead{Description} & \colhead{$R_{b,0}$} & \colhead{$\alpha$} 
}
\startdata
AD53 & Adiabatic expansion of a preformed hydrodynamic bubble with $\Gamma=5/3$ & $\left({3E_\Gamma \over 4\pi \eta p_0}\right)^{1/3}$
 & $3\beta \over 10$\\
AD43 & Adiabatic expansion of a preformed hydrodynamic bubble with $\Gamma=4/3$ & $\left({3E_\Gamma \over 4\pi \eta p_0}\right)^{1/3}$ & $3\beta \over 8$\\
CIH & Continuous inflation of a hydrodynamic bubble with $\Gamma=4/3$ & $\left({3 \dot{E} t \over 4\pi \eta p_0}\right)^{1/3}$ & $3\beta \over 8$\\
FML & Bubble threaded with flux-frozen magnetic loops & $\left( {{8\pi\,\mu_b^2}\over{p_0} } \right)^{1 \over 6}$
& $3\beta \over 8$\\
CDJ & Current-dominated magneto-hydrodynamic jet model & ${4\pi \over c}  \left(8\pi\,p_0\right)^{-{1 \over 2}} \, I_z $ & $3\beta \over 4$\\
\enddata
\end{deluxetable*}

%%%%%%%%%%%%%%%%%%%%%%%%%%%%%%%%%%

\section{Observations}\label{s.observations}

\subsection{Bubble Size vs. Radius}

\begin{figure}[t]
\begin{center}
\includegraphics[width=0.45\textwidth]{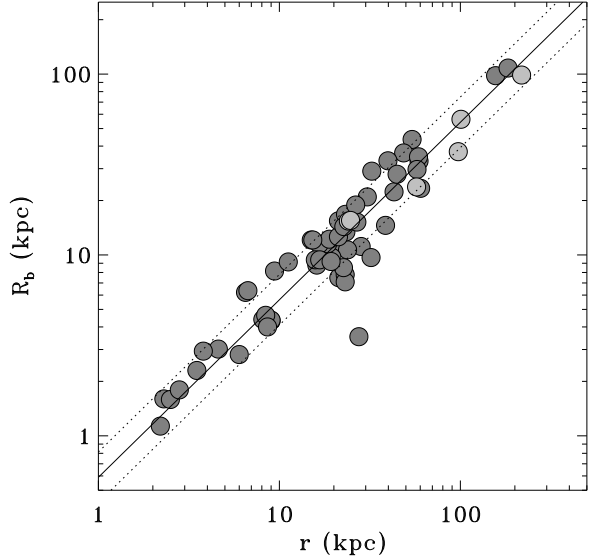}
\end{center}
\caption{Bubble size as a function of bubble location in physical units (kpc). The light grey data points indicate the data points for the multi-cavity system Hydra~A. The solid line indicates the best fit correlation, while the dashed line show the $1\sigma$ intrinsic width of $0.098,{\rm dex}$. \label{f.r_rb_kpc}}
\end{figure}

Taking our sample of 64 bubbles in 32 clusters, we can now analyze the evolution of the bubble size as a function of distance to the center in a statistical sense, to distinguish between the models discussed in the last section. Figure \ref{f.r_rb_kpc} shows the bubble size as a function of its physical projected distance to the center in $\kpc$. The solid line shows the best fit correlation, while the dashed lines indicate its $1\sigma$ intrinsic width. With an intrinsic width of only $0.098,{\rm dex}$, this correlation is surprisingly tight, considering the fact that our sample comprises clusters of different sizes, richness, black hole masses, etc, and even extends into the group and galaxy scale. We also emphasize that this correlation covers almost three orders in magnitude and is {\it not} simply a consequence of scaling both axes with distance. The correlation is as strong and as tight when plotted in terms of apparent angular sizes. We will discuss in section \ref{s.discussion} how incompleteness and other effects conspire to produce such a tight correlation. 

\begin{figure}[t]
\begin{center}
\includegraphics[width=0.45\textwidth]{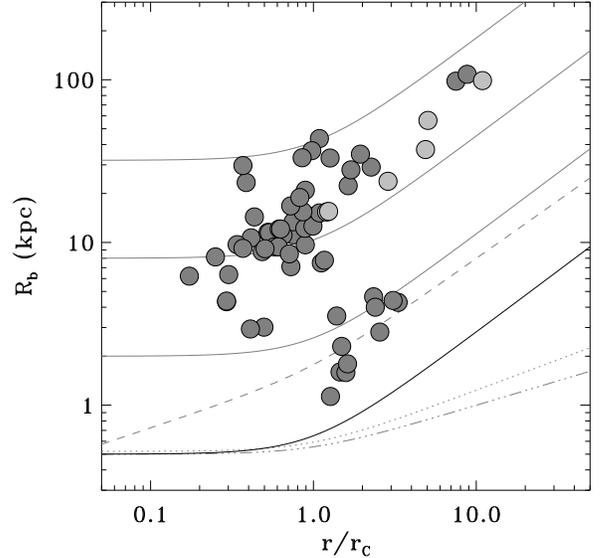}
\end{center}
\caption{Bubble size as a function of bubble location scaled by the core radius. The light grey data points indicate the data points for the multi-cavity system Hydra~A. The lines show the predictions from our models for one particular initial bubble size: magnetically dominated model (solid lines); purely hydrodynamic expansion with an adiabatic index of $\Gamma=5/3$ (triple-dot-dashed line) and $\Gamma=4/3$ (dotted line); and the flux-frozen magnetic dipole model (also dotted line). The grey solid lines show the magnetic model for different initial bubble sizes. Note that all lines show predictions for deprojected radii, whereas the data show the projected radius as measured in the images, as we do not have information about the jet geometry. Deprojection will systematically shift the data points to the right. Note that the shape and size of the shown error ellipse should be also a fair representation of the uncertainty in the fitted parameters themselves (red star).
\label{f.r_rb}}
\end{figure}

As all model predictions scale with $[ 1+ (r/r_c)^2]^{\alpha}$, it is more instructive to scale the bubble's distance to the center by the core radius, as shown in Figure~\ref{f.r_rb}. The triple-dot-dashed line shows the shallowest evolution of the AD53 model. The dotted line indicates both the FML model, as well as the AD43 model. The steepest model predictions shown are the magnetically dominated bubbles inflated by current-carrying jets (CDJ, solid line) and the continuously inflated hydrodynamic model (CIH, dashed line). This diagram is a rather powerful probe, as it distinguishes between the power of the outbursts that have supposedly created the bubbles, and their subsequent evolution. However, note that the exponent $\alpha$ also depends on the $\beta$ parameter, which is centered around $0.50$ for our sample (see Table \ref{t.clusterprop}). All lines drawn in our Figures use this average value.

Figure \ref{f.r_rb} demonstrates how the AD53, AD43 and FML models are all too shallow to represent the data accurately, while the CDJ model and the CIH model adequately reproduce the steep slope of the correlation. To quantitatively support this impression we produce Monte-Carlo simulations of the various models in section~\ref{s.discussion}.

The rather wide spread in the correlation over several orders of magnitude in bubble size in Figure~\ref{f.r_rb} must be due to the a difference in initial bubble sizes or energy injection rates. In each of the presented models, the initial bubble size and in turn the cavities $pV$ work is related to the central pressure $p_0$ and the total AGN outburst energy, either directly (eq. (\ref{e.rb_gamma})), through means of the magnetic dipole (eq. (\ref{e.rb_dipole})) or the current (eq. (\ref{e.rb_current})). Thus, any spread perpendicular to the indicated lines for any model predictions in this figure is primarily due to the spread in the initial bubble size, and consequently AGN power.

\begin{figure*}[h]
\begin{center}
\includegraphics[height=0.29\textwidth]{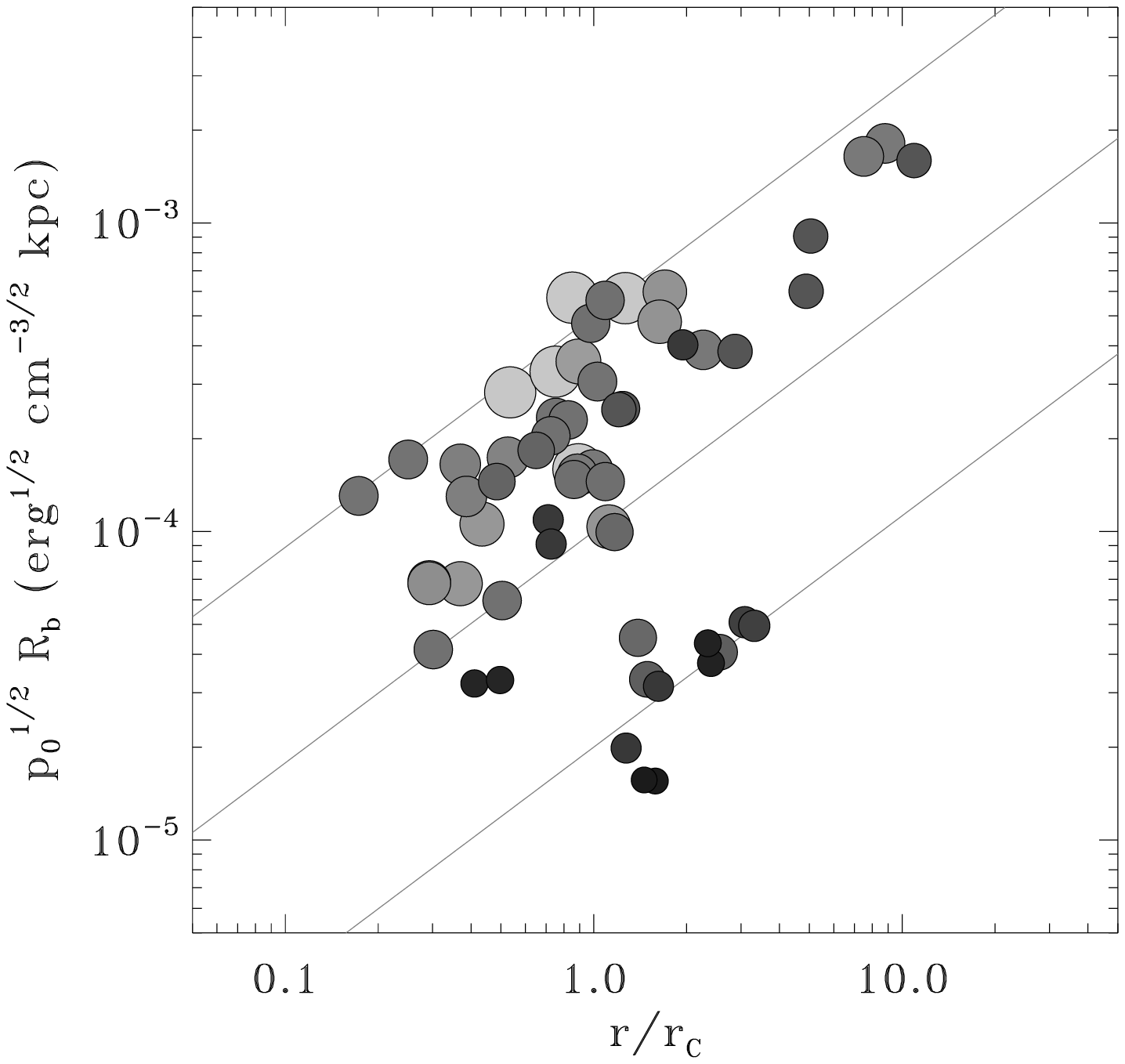}\hspace{-.4cm}
\includegraphics[height=0.29\textwidth]{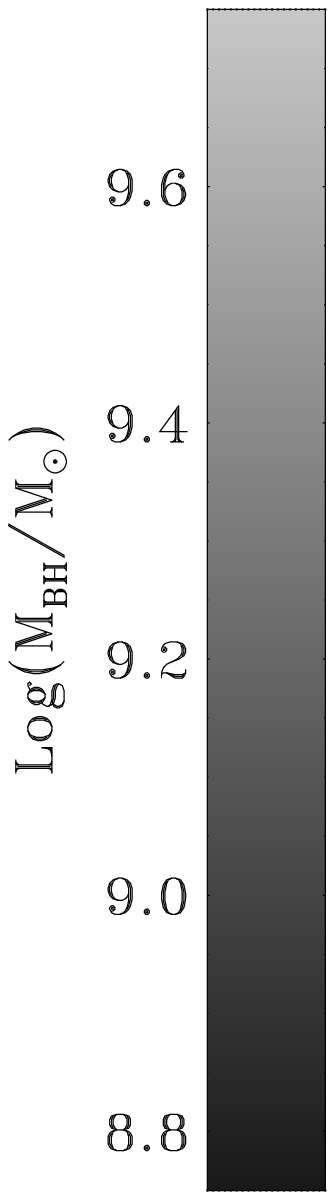}\hspace{.4cm}
\includegraphics[height=0.29\textwidth]{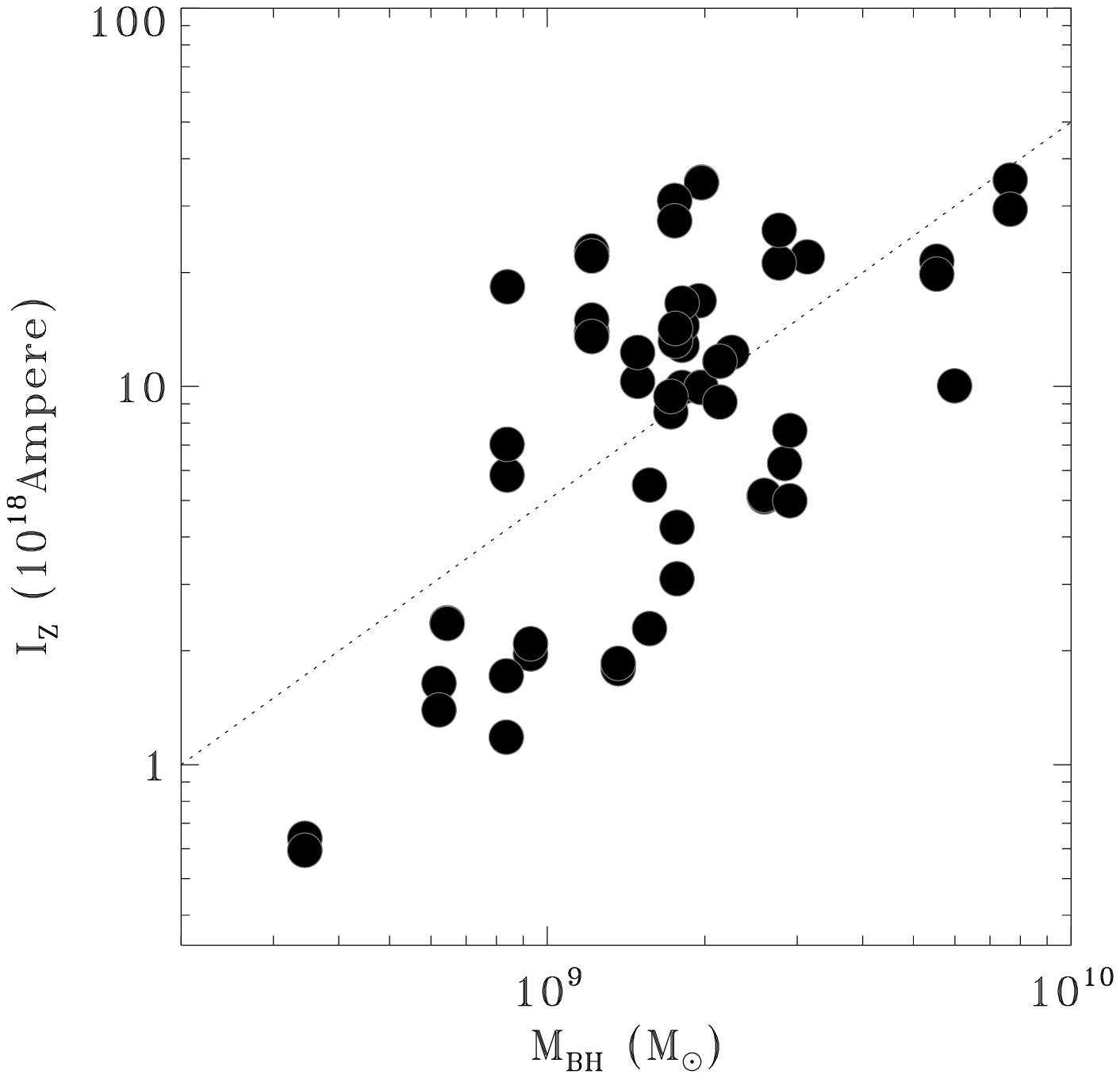}
\includegraphics[height=0.29\textwidth]{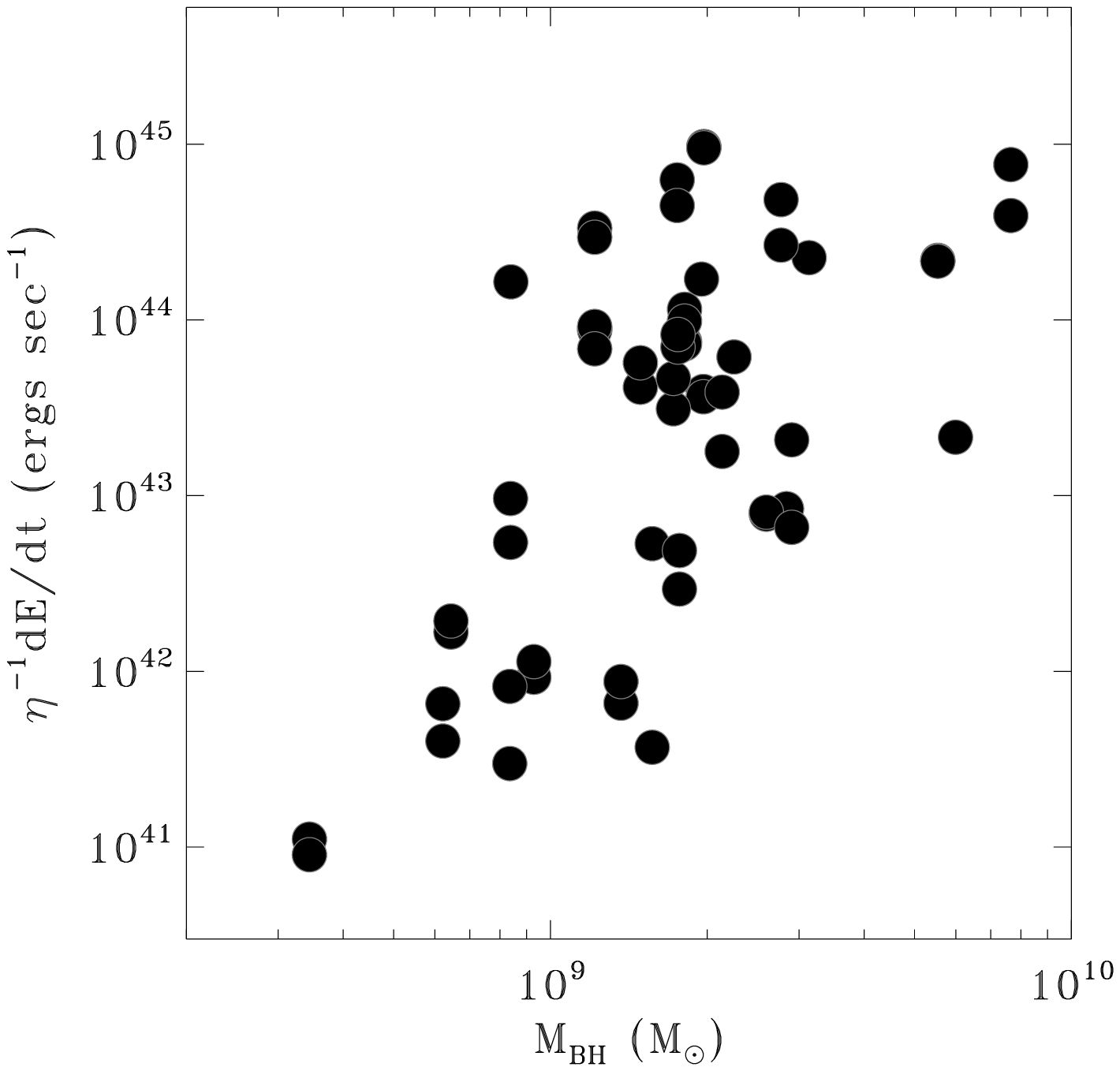}\\
\includegraphics[height=0.29\textwidth]{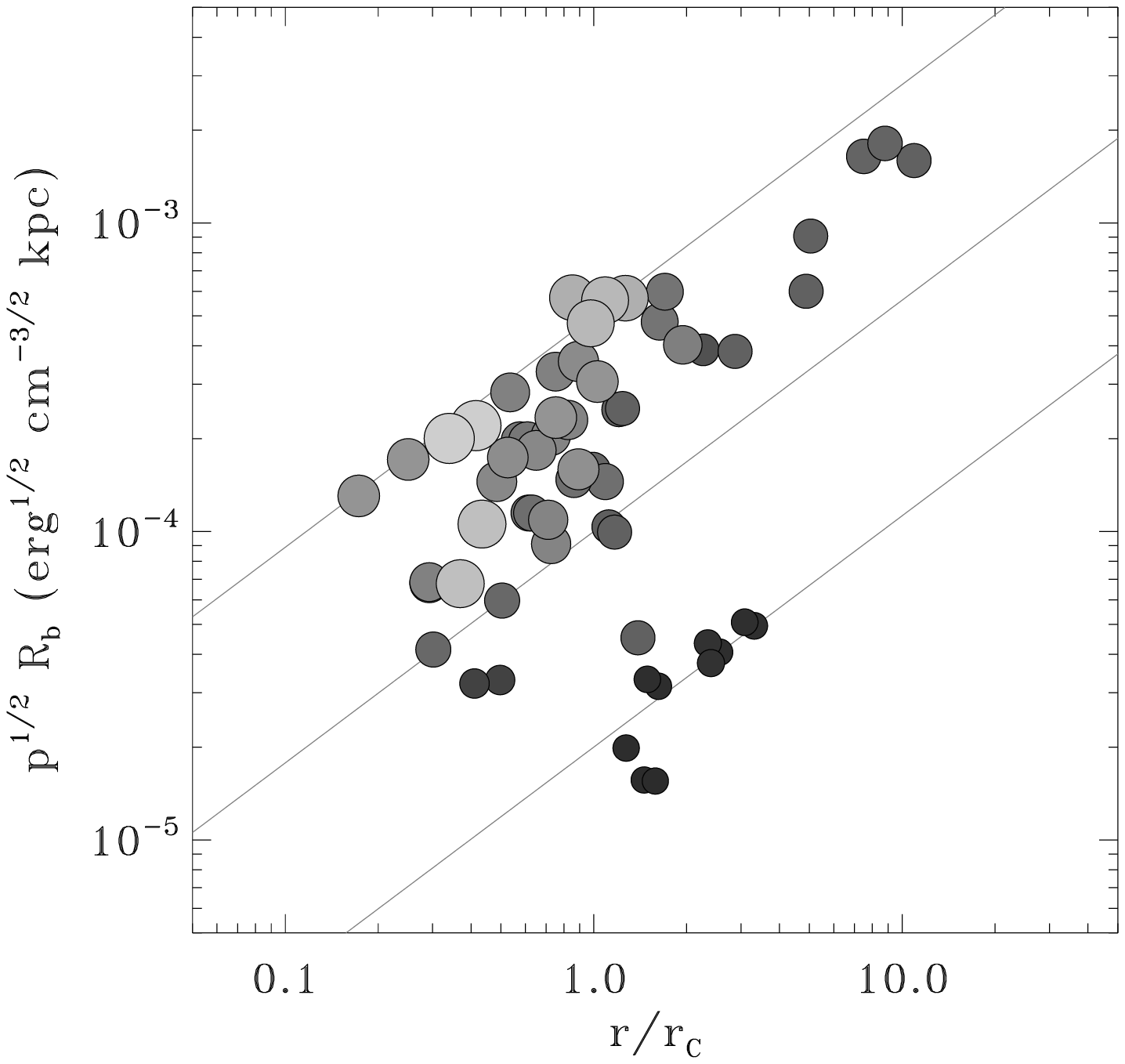}\hspace{-.4cm}
\includegraphics[height=0.29\textwidth]{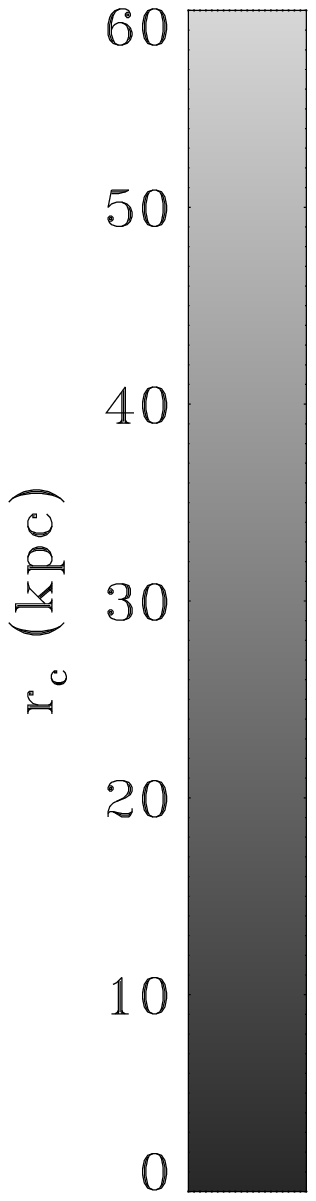}\hspace{.4cm}
\includegraphics[height=0.29\textwidth]{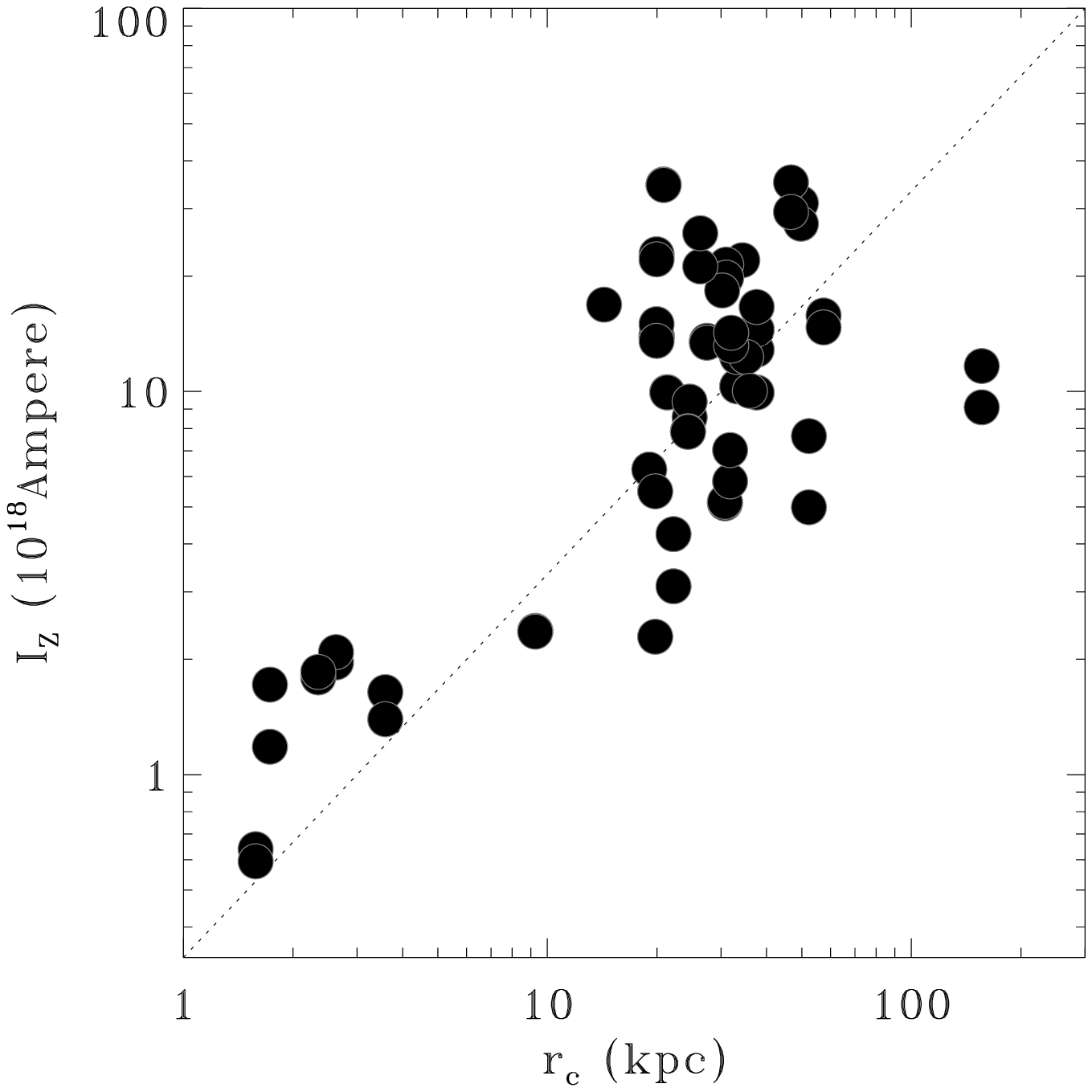}
\includegraphics[height=0.29\textwidth]{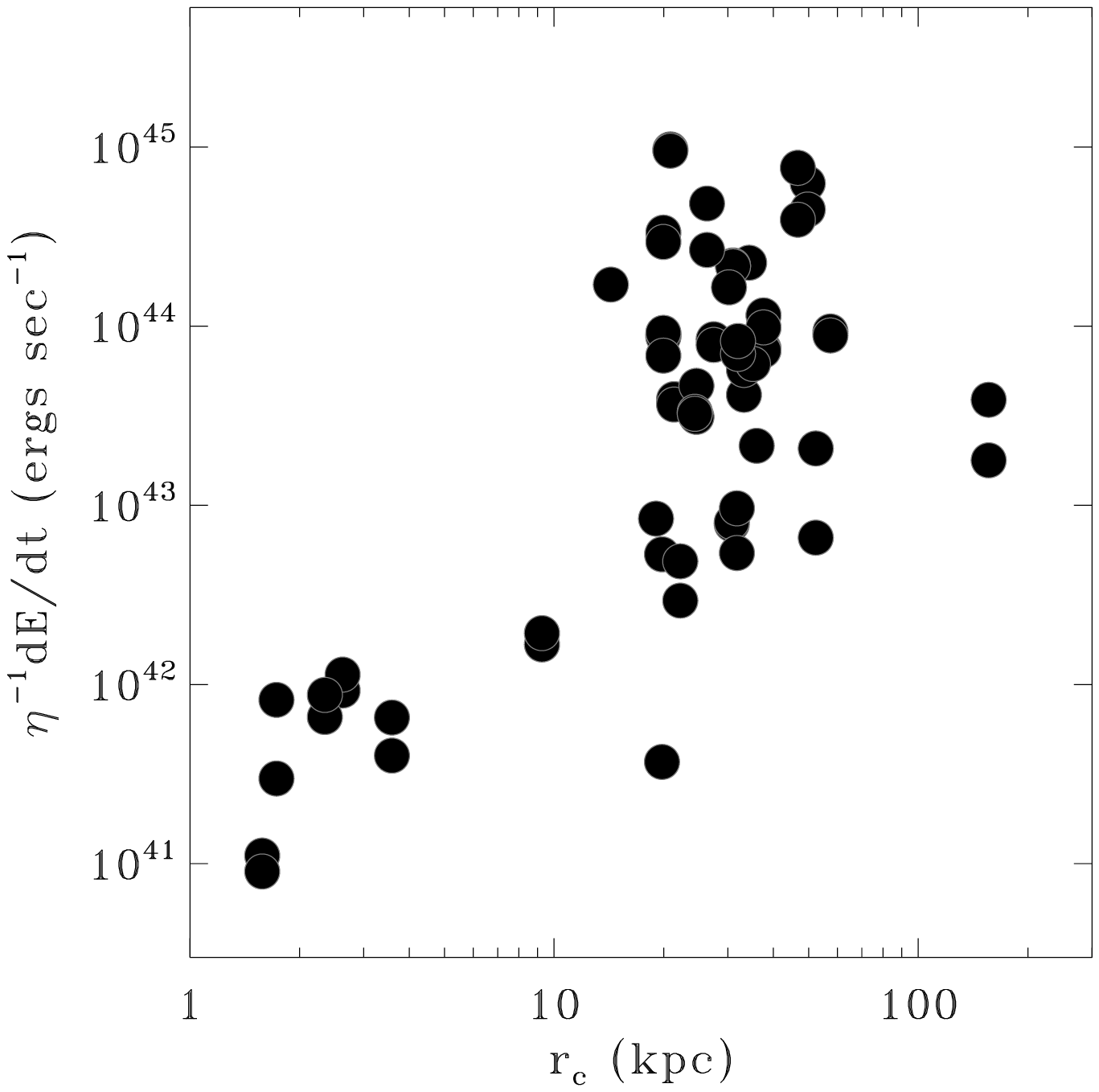}\\
\includegraphics[height=0.29\textwidth]{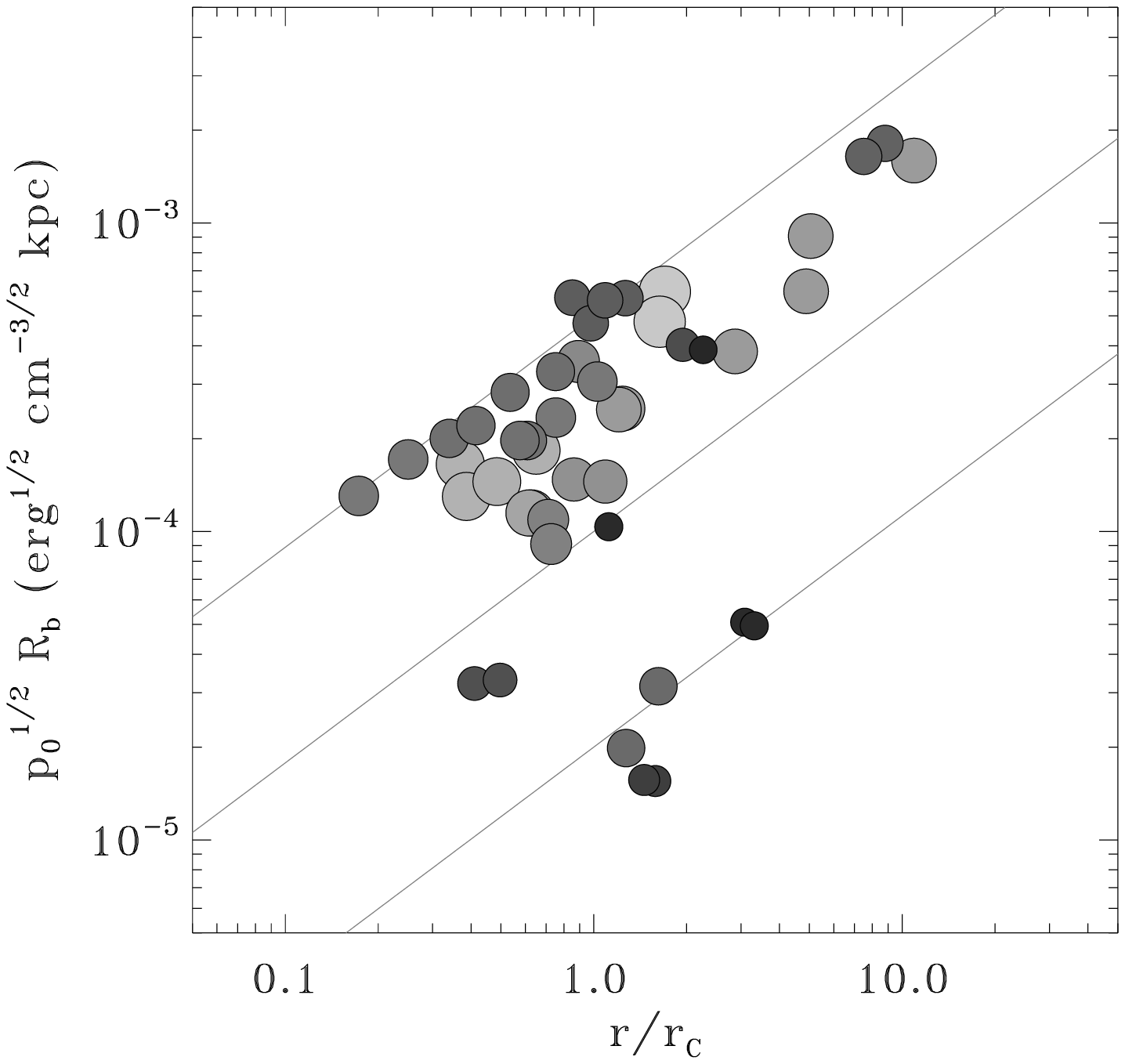}\hspace{-.4cm}
\includegraphics[height=0.29\textwidth]{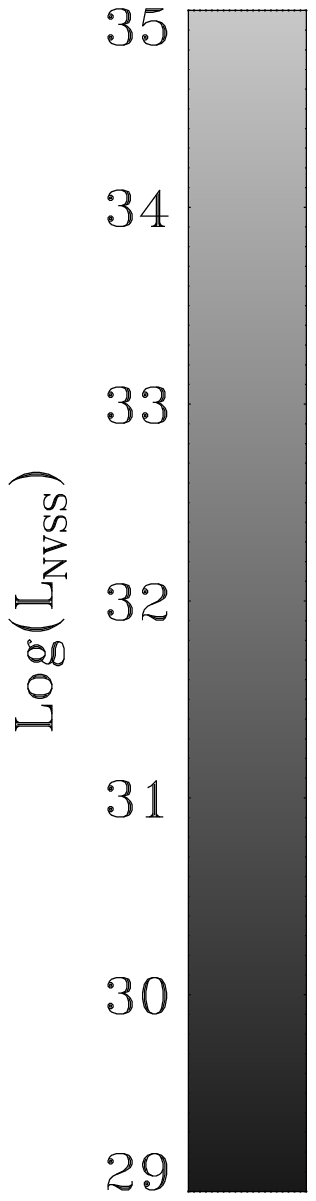}\hspace{.4cm}
\includegraphics[height=0.29\textwidth]{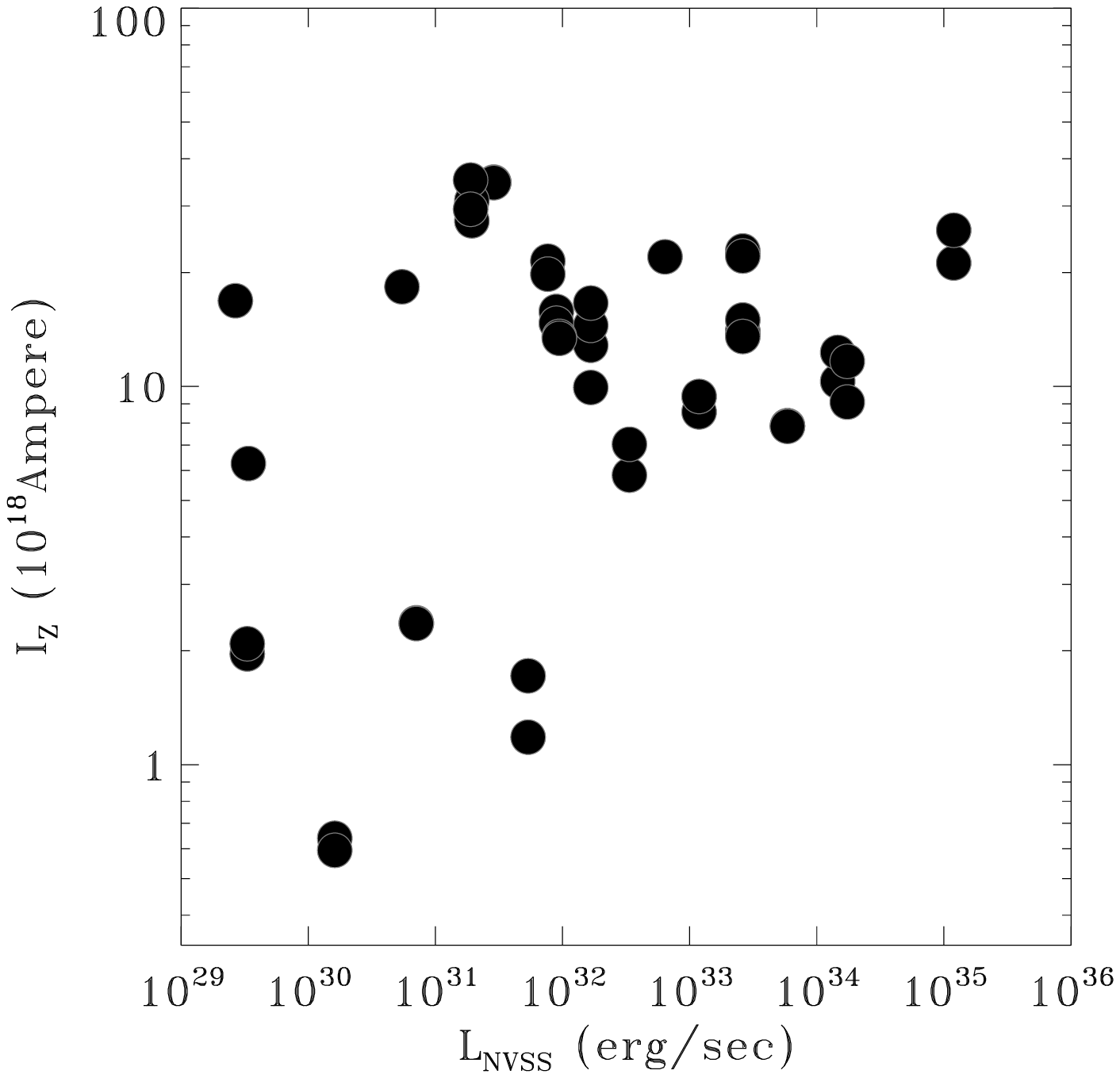}
\includegraphics[height=0.29\textwidth]{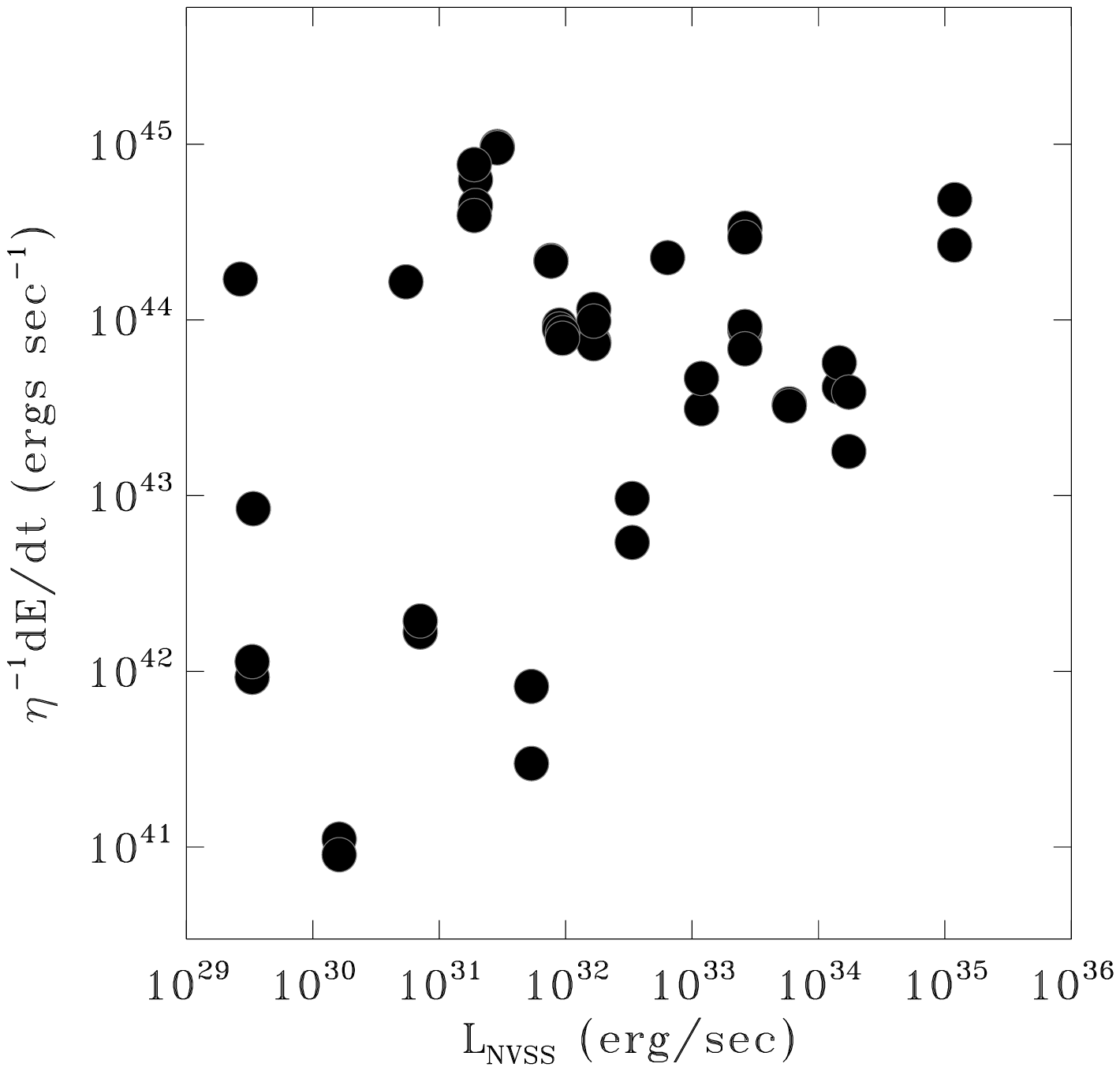}
\end{center}
\caption{The left column shows the scaled bubble radius as a function of bubble location. Colors and symbol sizes indicate the dependence on independent third parameters, namely black hole mass (top), core radius (middle) and NVSS radio luminosity (bottom). Note how they tend to band the plot along the indicated guide lines, representing the approximate asymptotic slope for the CDJ and CIH models. The middle panel plots these same quantities against the inferred current for the CDJ model. The dashed line indicates a simple proportionality between the parameters. Both black hole mass and core radius scale almost linearly with current, though with significant scatter. The right panels show the equivalent energy injection rate for the CIH model. Please note that both the current and the energy injection rate are derived from the cavity properties on the assumption of a face-on geometry. Thus, projection issues account for some of the scatter in these plots. \label{f.Iz_mbh_rc} \label{f.r_rb_mbhrccolor}}
\end{figure*}

To test whether the location in the $r/r_c$ vs. $R_b$ diagram in Figure \ref{f.r_rb} is correlated with any other cluster characteristic, we have produced color-coded plots with different parameters. Two strong correlations emerge, which are shown in two top left panels of Figure \ref{f.r_rb_mbhrccolor}. In both plots, we have scaled the bubble size by the square root of the central pressure, according to equation (\ref{e.r0_current}) for the current-dominated bubbles. While this does not make a large difference for the general appearance of the plot, the y-offset is now solely due to the current itself for the CDJ model. Scaling the bubble size with $p_0^{-1/3}$ or $p_0^{-1/6}$ as needed by the other models, shows qualitatively the same trends. The top left panel of Figure~\ref{f.r_rb_mbhrccolor} marks the data points by central black hole masses, in the sense that larger symbols with lighter colors denote more massive black holes. Black hole masses are derived from velocity dispersion ($\sigma$) data from Hyperleda \citep{Hyperleda} using the relation given by \citet{TremaineBHsigma}. Thus, this apparent separation of the plot with black hole mass could also be rather a separation with $\sigma$ itself. The middle left panel of Figure \ref{f.r_rb_mbhrccolor} shows an even stronger trend with the core radius. Here, the larger symbols and lighter colors indicate larger core radii. We will quantitatively address these trends in section \ref{s.montecarlo} later on. 

In both plots, the data points are mostly separated perpendicular to the guidance lines drawn in the plot. These guidance lines show a slope of $3/4$, the asymptotic slope for the CDJ model for a $\beta$ of $0.5$, which is the average $\beta$ in our sample. The asymptotic slope of the CIH model is $17/24$, which is only $1/24$ shallower than the CDJ prediction. Note that the slope for the next steepest correlation ($3/8$) is already half as steep as the plotted lines, clearly at odds with these models. With a larger cavity sample being available in the future, this type of study will greatly benefit from splitting the sample into appropriate sub-samples that cover only small parameter spaces to find out what set of parameters truly governs the AGN power. 

As the bubble size have been scaled by the square root of the central pressure, the location of the various cavities in the plot are solely determined by the current flowing through the jet-lobe system -- assuming the CDJ model is indeed correct. In the middle column of Figure~\ref{f.Iz_mbh_rc}, we thus plot the inferred current $I_z$ as a function of black hole mass and core radius. One caveat to keep in mind is that $I_z$ is derived under the assumption that the geometry of the observed system is face-on. This projection effect will add an uncertainty to the derived current, and in general our reprojected bubble sizes and thus our currents will be slightly overestimated as a result, on average by a factor of $1.6$. The projection effect will also add a significant amount of scatter into correlations with $I_z$, such that 90\% of all data points would lie within a factor of $2.4$. In both plots, the correlation that we find is almost linear, as indicated by the dotted guidance lines. We find that the inferred currents span about two orders of magnitude, ranging from $\sim 5\times 10^{17}$ Amp\`ere to $\sim 5\times 10^{19}$ Amp\`ere. 

The right panels of Figure \ref{f.r_rb} show the same analysis for the CIH model, but for the energy injection rate instead of the currents. The panels show similar trends, with only slightly more scatter than for the CDJ model.

The bottom panel of Figure \ref{f.r_rb} on the other hand shows the same correlations for the integrated $20\cm$ radio luminosity within $1\arcmin$ from the NVSS catalog \citep{NVSS}. The observed trend is much weaker, as can also be seen in the middle and right panels. There appears to be a large scatter in this correlation, but an apparent lack of clusters with high radio luminosity but low current (or low energy injection rate) inferred from the cavity sizes. This is interesting, as radio luminosity is quite often used as a surrogate for total AGN power. This work may suggest otherwise, at least at the wavelength and the surface brightness level of the rather shallow depth of the NVSS survey. Radio observations at longer wavelengths may be better suited for this task \citep{birzan2007}.

\subsection{Bubble energy increases with Radius}

\begin{figure}[h]
\begin{center}
\includegraphics[width=0.45\textwidth]{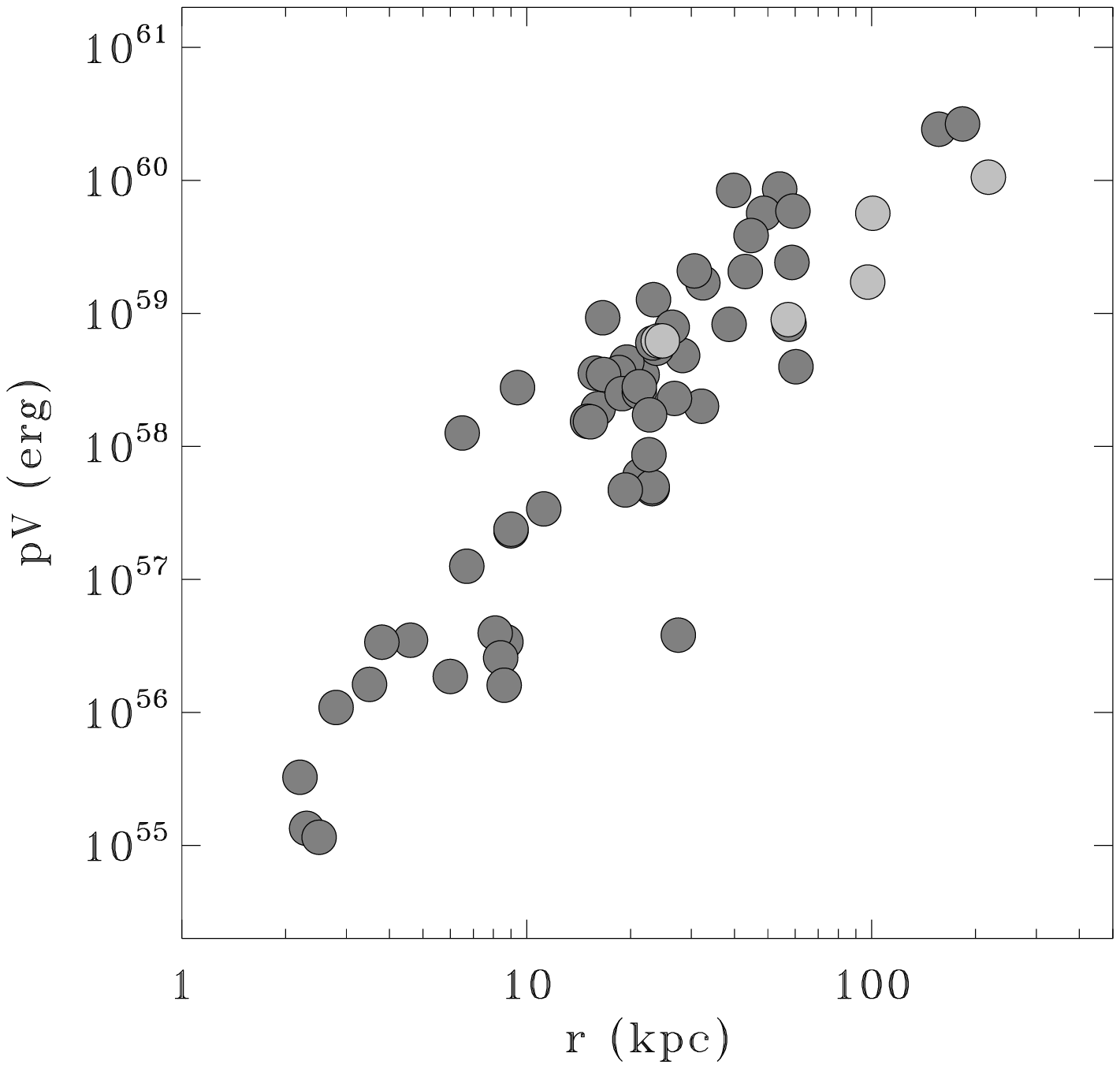}\\
\includegraphics[width=0.45\textwidth]{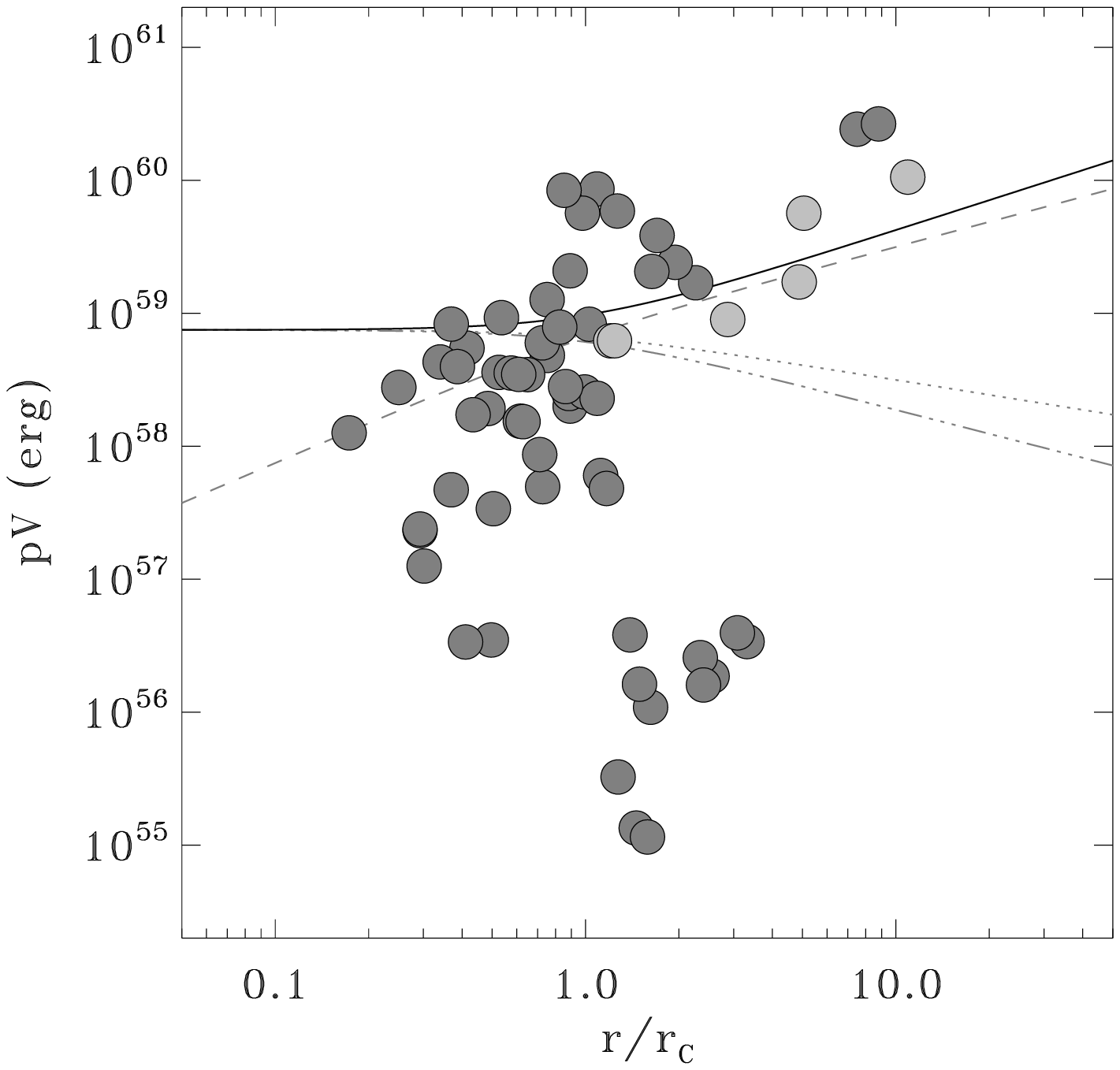}
\end{center}
\caption{{\it Top panel}: Bubble $pV$ energy as a function of physical radius. Note that there is a trend of bubbles in the outskirts of clusters to have higher energies. The light grey shaded data points show the multi-cavity system Hydra~A. {\it Bottom panel}: Same plot but with the radius scaled by the core radius of the cluster. The lines show the predictions from our models: the CDJ (solid line), CIH (dashed line), AD53 (triple-dot-dashed line) and AD43 (dotted line), and the FML model (also dotted line). Note that only the CDJ and CIH models actually predict an inferred increase in $pV$ energy with radius, all other models predict a declining behavior.\label{f.pV}}
\end{figure}

It is nowadays standard procedure to measure the size of a bubble and the ambient pressure at its location to get an estimate of the $pV$ work needed to inflate the bubble, which is then in turn used to infer the energy associated with the AGN outburst. Interestingly, if one plots the inferred energy as a function of radius for the whole sample, one finds that the energy increases as a function of bubble distance from the center, as shown in Figure \ref{f.pV}. This trend is obvious in the top panel figure, showing $pV$ as a function of the physical radius (in kpc). However, there is really no {\it physical} reason for bubbles at a larger radius to be produced by outbursts of higher energy, the radius should simply depend on the rise velocity, and the time elapsed since the outburst occured. 

\begin{figure}[h]
\begin{center}
\includegraphics[width=0.45\textwidth]{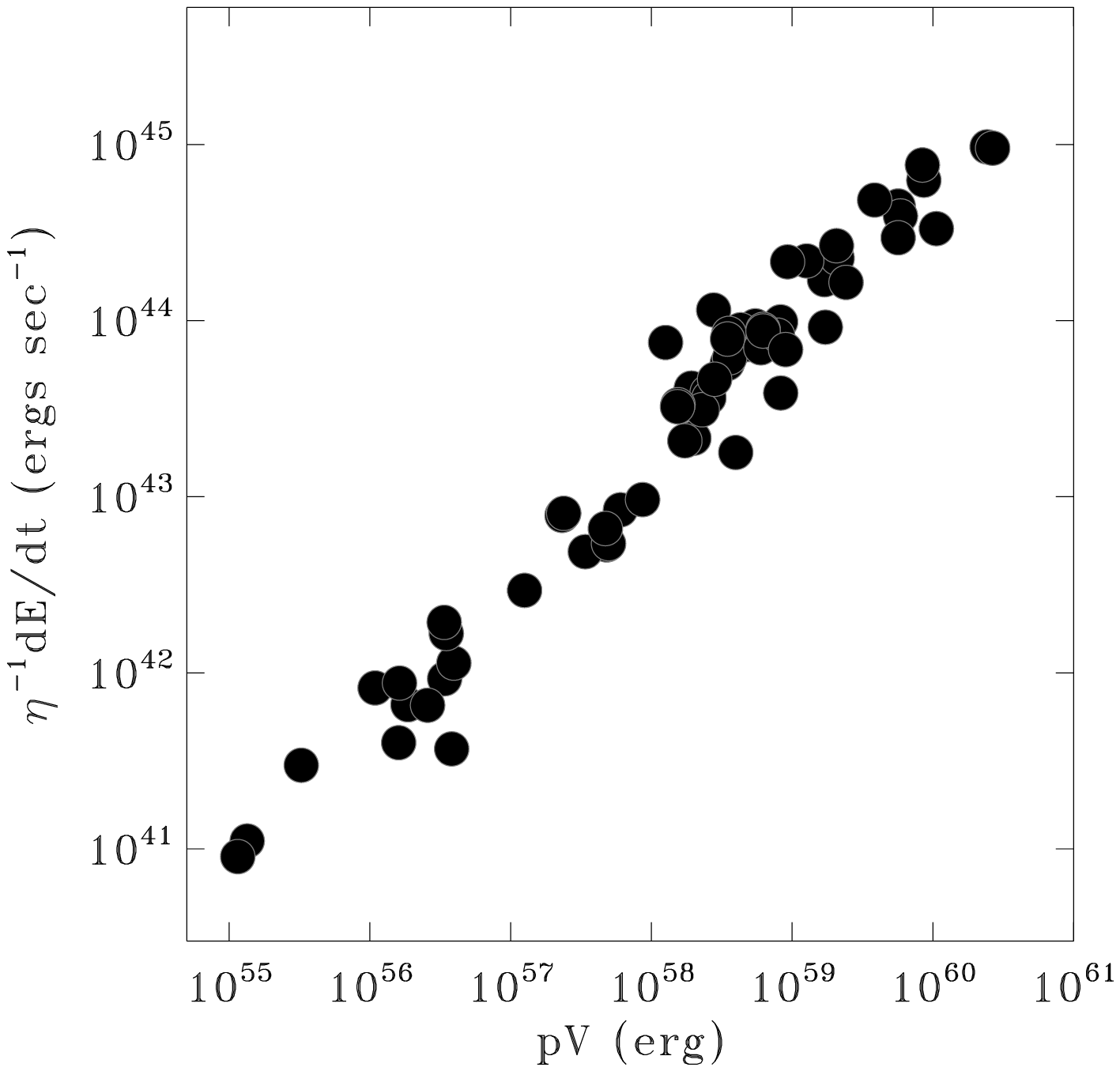}\\
\includegraphics[width=0.45\textwidth]{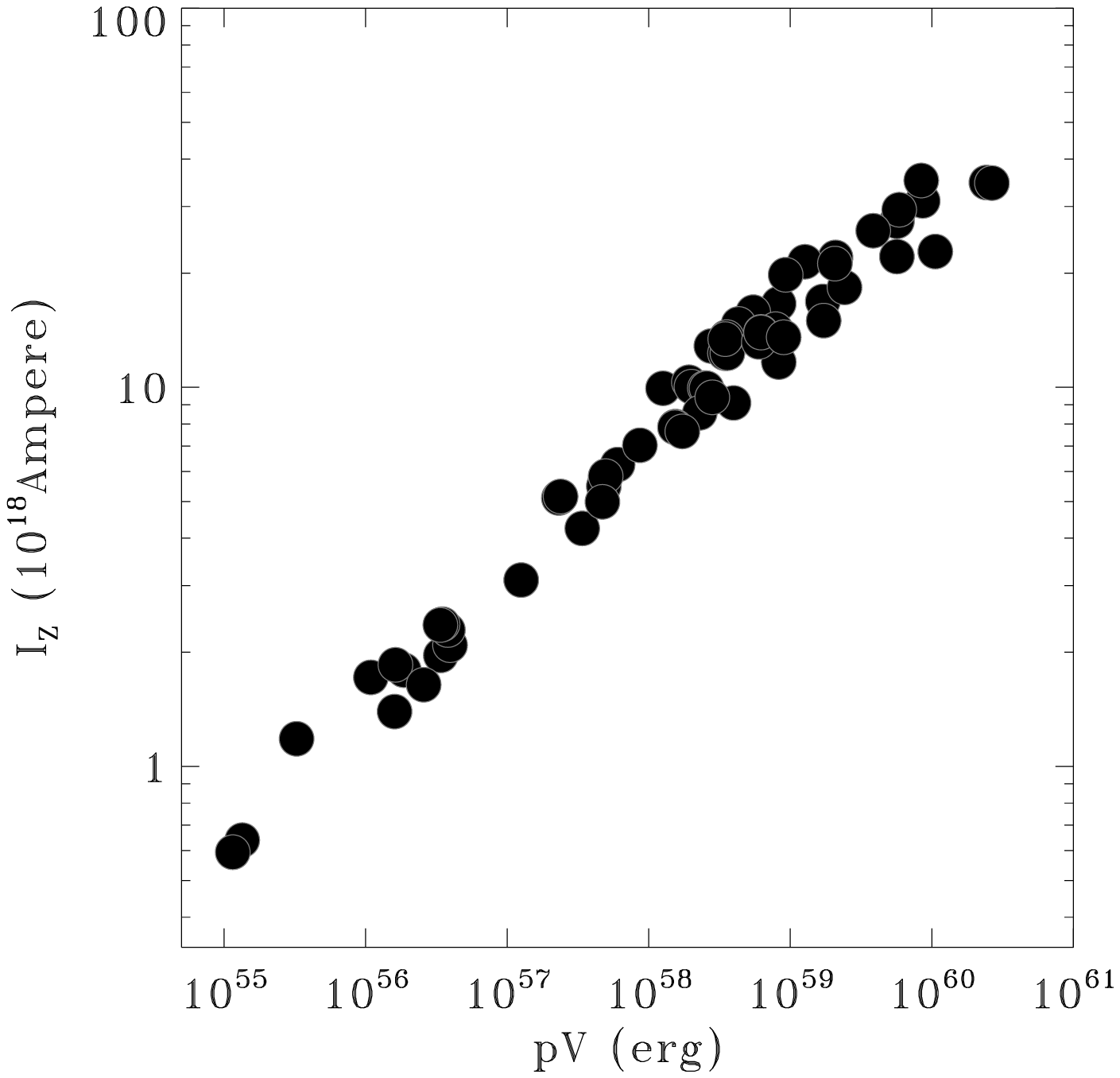}
\end{center}
\caption{Bubble $pV$ energy as a function of inferred energy injection rate for the CIH model (top panel) and current for the CDJ model (bottom panel). Note that in each plot, both parameters are derived from common parameters, which may account for some of the tightness in the correlation. However, the correlation with the current is significantly tighter than with energy injection rate. The remaining scatter is consistent with projection effects alone. \label{f.pVIz}}
\end{figure}

The bottom panel in Figure \ref{f.pV} shows the same data points, but this time with the bubble location again scaled by the core radius as before. This allows us to compare the data to predictions of the various models. Since we know the pressure profile as a function of radius, and we know the bubble size as a function of radius for a given outburst energy, we can easily calculate the observable $pV$ work as a function of radius. In fact, three out of the five models predict that the inferred $pV$ value should {\it decrease} with radius. This is due to the fact that in the adiabatic model (AD43: dotted line, AD53: triple-dot-dashed line) part of the internal energy needed to keep up the internal pressure support of the bubble is used to expand the bubble adiabatically. In the FML moment model (also dotted line), this energy is used up to expand the magnetic flux loops on the bubble surface. All of these models are again inconsistent with the data, along the same lines of argument presented in the last section. The only models in which $pV$ is expected to increase with radius are the CDJ and CIH models. Figure \ref{f.pVIz} shows the correlations between the inferred $pV$ and the $I_z$ and $\dot{E_0}$ values for our sample. Note how a spread of only two orders of magnitude in current results in $pV$ values spanning almost six orders of magnitude. Part of the reason for the intrinsic tightness of these correlations is that the parameters are all derived from a common set of measured variables (see eq. \ref{e.r0_current} and \ref{e.rb_inflated}).

We emphasize that this discussion should not be considered an additional argument for the CDJ or CIH models, as the $pV$ values are strongly correlated with the bubble sizes. We rather want to make the reader aware of the severe implications this has on the inferred energetics of AGN feedback. It is now accepted standard procedure to use the adiabatic model and $pV$ derived outburst energies to determine the overall energy budget of the clusters, and to find out whether AGN heating can offset cooling in clusters \citep[e.g.][]{BirzanCavities, BestAGNcooling}. However, Figure \ref{f.pV} makes absolutely clear that we have a very biased view of bubbles in clusters. If the adiabatic model can really be considered a viable model, one needs a way out of this apparent paradox and understand why outbursts seen at larger radii tend to have larger $pV$ values. This effect is not only apparent in our large sample of different clusters, but even within the few individual clusters with known multiple bubbles, such as Hydra~A and Perseus. In Hydra~A, \citet{WiseHydraA} have found a multi-cavity system with three bubble pairs at different radii and inferred $pV$ energies that continuously increased as a function of radius, with the inferred outburst energy several times higher for the outermost bubbles. Recently, \citet{SandersPerseusNewcav} reported the possible detection of another cavity in Perseus at very large radii ($\sim 170\kpc$), and also inferred an outburst energy several times higher than the latest outburst.

There are a few ways out of this dilemma. The first possibility is that we simply need a more complex way to infer outburst energy from $pV$, and our current picture is too simplistic and thus incorrect. In this case, theoretical models have to step up and provide better constraints for X-ray observables to see if our simple methods actually do have a rigorous basis to stand on. 

The other way out is to assume that nearly all of our bubbles undergo {\it continuous} inflation, even at this moment. However, a large fraction of our bubbles are considered ``ghost bubbles'' and seem to be no longer directly attached to the jet or associated with young radio emission. On the other hand, many cavities that were earlier categorized as ``ghosts'' now show evidence for older low-frequency emission \citep[e.g.][]{ClarkeXrayTunnel}, and in some cases an X-ray tunnel of suppressed surface brightness may still exist that connects to cavities further out \citep[e.g.][]{ClarkeXrayTunnel, WiseHydraA}. Creating a purely hydrodynamic jet model capable of continuously inflating these bubbles at large radii is challenging and -- in our opinion -- unlikely to succeed. In particular, the geometric configuration of the cavity system in Hydra~A (Figure \ref{f.HydraA}, left panel) would be difficult to explain with a conventional hydrodynamic jet model, due to its apparent ``kink''  before the outermost cavity pair. A magnetic jet on the other hand could be deformed by instabilities or ambient gas flow and still be able to deliver energy along the field lines to inflate the bubbles even at large radii.

The third and last solution to this paradox is the effect of incompleteness, that we are going to address in depth in the discussion section of this paper. Understanding this effect will be crucial in correctly addressing the question whether AGN can be a general solution to the cooling flow problem in clusters and galaxies. To do this, we also have to properly understand the size evolution of the bubbles, as these two effects are of course intimately connected.

\subsection{The Multi-Cavity System Hydra~A}

Understanding the evolution of bubble sizes in a large sample of clusters gets complicated by the fact that the sample is a combination of a wide variety of cluster properties. A much cleaner approach is to look at the evolution of bubbles in a {\it single} system. Since we are of course unable to observe a single cavity multiple times during its rise in the cluster due to the extremely long rise times (several Myr), we have to rely on systems that harbor multiple cavities at different distances. Based on the reasonable assumption that AGN produce at least similar, if not identical bubbles over time, we can then use these individual bubbles to infer the size evolution of a rising bubble over a range of distances. 
 
\begin{figure*}[t]
\begin{center}
\includegraphics[width=0.45\textwidth]{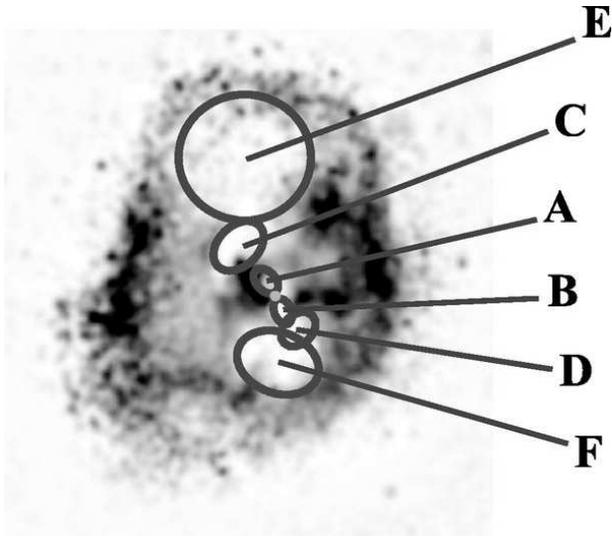}
\hspace{0.05\textwidth}
\includegraphics[width=0.45\textwidth]{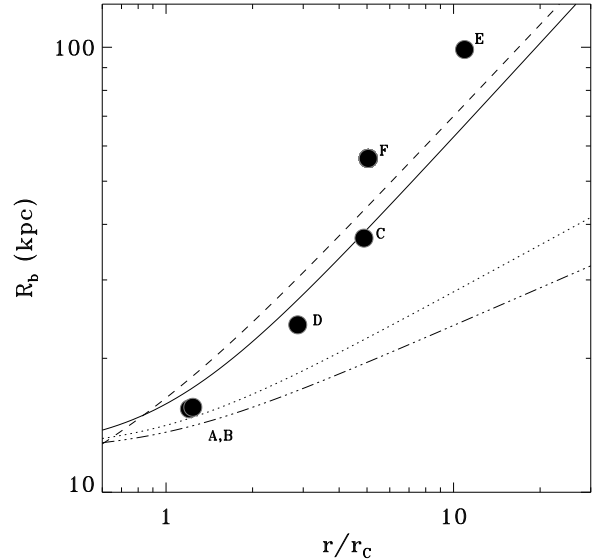}
\end{center}
\caption{
{\it Left:} The multi-cavity system in Hydra A, reproduced from \citet{WiseHydraA} with permission from the authors. The black area is excess X-ray emission left-over after an elliptical surface brightness model has been subtracted. 
{\it Right:} Data Points: Bubble sizes for Hydra A as a function of distance to the center, taken from \citet{WiseHydraA}; Lines show predictions from the AD53 (triple-dot dashed line), AD43 (dotted line), FML (also dotted line), CIH (dashed line), as well as the CDJ model (solid line). The cavity labels are the same in both plots. \label{f.HydraA}}
\end{figure*}

In the case of Hydra~A, \citet{WiseHydraA} lately discovered a giant multi-cavity systems consisting of 6 distinct cavities, with the outermost bubble being intact at a distance of over $200\kpc$ from the center. The left panel of Figure \ref{f.HydraA} shows the cavity system, reproduced from \citet{WiseHydraA} with permission from the authors. The properties of these bubbles are also listed in Table~\ref{t.bubbleprop}. 

Since Hydra~A is now the best multi-cavity system with adequate published data to test our models, we have also marked its data points with light grey in Figures \ref{f.r_rb_kpc}, \ref{f.r_rb} and \ref{f.pV}. The right panel of Figure \ref{f.HydraA} shows the cavity sizes as a function of distance from the center for only the Hydra~A cavities. Note how quickly the cavities expand as they rise outward. They tend to grow much faster than expected for the purely hydrodynamic models AD53 and AD43 (triple-dot-dashed and dotted lines, respectively) and the magnetic dipole model (FML, also dotted line). The only two models in agreement with all six cavity sizes are the CDJ and CIH models again. If the CDJ model is indeed correct, the jet in Hydra~A carries a current of $\sim 1.5-2\times10^{19}\, {\rm A}$, for the CIH model, this would require an average energy injection rate of $\sim 0.8-3\times10^{44} \erg \sec^{-1}$ over the largest cavity life time of $\sim 230\,{\rm Myrs}$.

A look at the radio emission of Hydra~A shows that these cavities all seem to be connected at low radio frequencies out to a few hundred kiloparsec \citep{TaylorHydraA}. There also seems to be ``kink'' in the radio emission close to the location of cavities $C$ and $D$ (Figure \ref{f.HydraA}). This may suggest that the outer cavity pair has a different orientation than the other two pairs, and that it has a larger inclination angle with respect to the plane of sky. This is consistent with the data points for $C$ and $D$ in the right panel of Figure \ref{f.HydraA} to be shifted towards smaller radii due to projection effects. This would also explain why the inferred current (or energy injection rate) for the outermost bubble pair is slightly larger than for the two inner bubble sets.

For projection effects to cause individual disconnected, purely hydrodynamic bubbles to produce this steep plot, the jet would have to conspire and be significantly closer to the line of sight with each successive cavity pair. While this is sufficiently unlikely in itself, the outermost bubbles would also have to be several times further away when deprojected, at which point the contrast in the bubble would be way too small to be detectable due to the drop in surface brightness at larger radii.

\subsection{The Multi-Cavity System Perseus}

The other well-known multi-cavity systems is the famous Perseus clusters \citep[e.g.][]{SandersPerseusNewcav}. Judging from the appearance of the inner set of bubbles in Perseus \citep{Perseus_holes}, the geometry of the cavity set right next to the cluster center seems to be rather complicated, and different from that of the outer set, both of which are part of our current data set. 

Only recently, \citet{SandersPerseusNewcav} noted a possible intact cavity on the outskirts of the Perseus cluster around $170\kpc$. This tentative cavity candidate has a width of approximately $73\kpc$ ($200\arcsec$, Jeremy Sanders, private communication). Unfortunately, the bubble is located off the main ACIS chip, and is thus not included in all exposures. In addition, the {\it Chandra} point spread function is rather large at this location, and the surface brightness level very low, which makes its detection difficult. As a result, the height of the possible cavity remains very uncertain, and can be anywhere from around $9\kpc$ ($25\arcsec$) to $25\kpc$ ($65\arcsec$), depending on the surface brightness model that one chooses to subtract. 

\begin{figure}
\begin{center}
\includegraphics[width=0.45\textwidth]{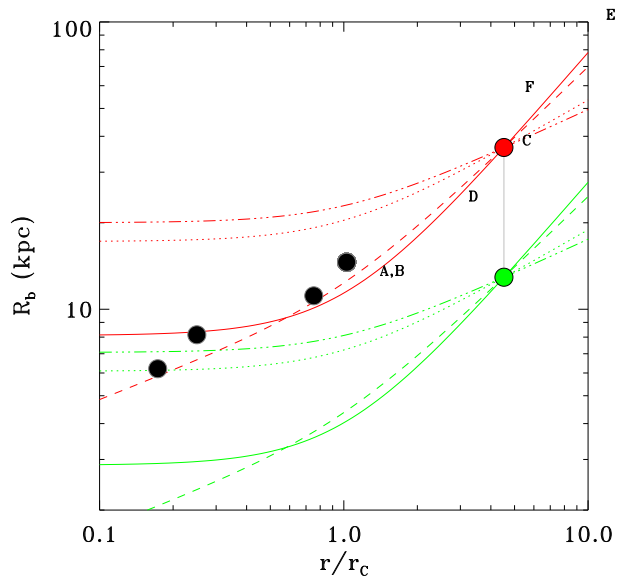}
\end{center}
\caption{Bubble sizes for Perseus as a function of distance to the center. Lines as in Figure \ref{f.HydraA}. The red data point  shows the upper limit for the new bubble size estimate, the green data shows a lower limit. The correct answer will likely lie somewhere in between these two extremes. \label{f.Perseus}}
\end{figure}

\citet{SandersPerseusNewcav} already note that the outburst energy associated with this new bubble requires an outburst of at least twice of that needed to create the inner bubbles. As we have pointed out in the last section, this could also be possibly explained by the CDJ or CIH models. Figure \ref{f.Perseus} shows the size evolution of the 4 observed cavities in the Perseus cluster (black filled circle). For the newly detected cavity candidate, we show two estimates for its bubble size: The red circle shows the width as the bubble size, the green circle uses the geometric average between the width and the smallest height estimate. The lines flowing through these two data points indicate how the bubble would have evolved to its current size assuming our different models. Unfortunately, our results remains inconclusive. If we assume that the green data point is representative of the bubble size, the energetics of the outburst that created this bubble is comparable to the outburst that is currently inflating the innermost set of cavities. However, if we take the red data point as given, the adiabatic models yield an outburst roughly double as energetic. It is intriguing to note that both the CIH and CDJ models are consistent with the larger bubble size estimate. A deep observation pointed directly toward this outer bubble (to have a smaller {\it Chandra} point spread function) is most likely needed in order to firmly assess the true size of this bubble and to paint a better picture of the outburst history of Perseus. More deep observations of other multi-cavity systems are needed to further distinguish between these models.

\section{Discussion}\label{s.discussion}

\subsection{Projection Effects}

Projection has several effects on the observables that are crucial to our analysis. The most obvious effect is that we are measuring the projected distance to the center, rather than the true distance. This will systematically shift the data points toward smaller radii in all plots. In fact, all of our radii should be considered lower limits to the true location of the bubbles. This will not only affect the radii themselves, but also the point at which other quantities are evaluated at, like density, temperature and pressure. In general the temperature rises outward in these systems, thus the temperature at the location of the bubble is likely to be systematically underestimated. The density and ambient pressure on the other hand will always be overestimated. This also means that any rise times derived from using the projected radius rather than the true distance to the center will result in estimates for the rise times that are systematically too low. We also note that the smaller the observed radius is, the higher the probability that it is due to an effect caused by projection. 

\begin{figure}[t]
\begin{center}
\includegraphics[width=0.49\textwidth]{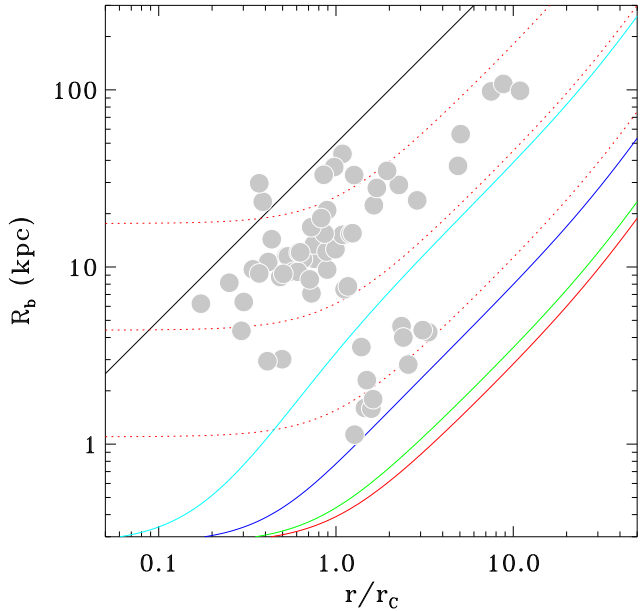}
\end{center}
\caption{Grey shaded data points are the same as in Figure \ref{f.r_rb}. The colored lines shows the required size evolution of a bubble to keep a constant signal to noise ratio while rising, taken from \citet{EnsslinBubbledetectability}, for different angles of inclination with the plane of sky: $0\degree$ (face-on, red), $30\degree$ (green), $60\degree$ (dark blue), and $80\degree$(light blue). The dotted red lines repeat the line for face-on detectability for different initial bubble sizes, spaced by a factor of 4. The black solid line indicates a slope of ``1'', to show equality between bubble size and projected radius, our ``geometric'' detection criterion. \label{f.r_rb_incompleteness}}
\end{figure}

But there are more subtle effects that projection has on our data. As we do not have an automated tool to detect bubbles, one has to rely on human experience in finding and identifying these systems. This task is much more difficult, if the cavities overlap with the bright cluster center or the bubble on the opposite side of the cluster. In fact, our sample does not contain {\it any} cavity system in which the bubble size exceeds the projected distance to the center, the slope of which is shown by the black solid line in Figure \ref{f.r_rb_incompleteness}, even though this is statistically very improbable. This suggests that our sample is affected by what we will refer to as a ``geometric'' selection effect, introduced by our manual detection process.

\subsection{Incompleteness Effects}\label{s.incompleteness}

One very important issue in a systematic study of X-ray cavities is to address incompleteness, i.e. how likely it is to detect a bubble of certain properties. \citet{EnsslinBubbledetectability} address this issue by computing analytical formulae for the $S/N$ ratio of a spherical hole in the isothermal $\beta$ model. In this context, the signal $S$ is the deficit in counts due to the hole, i.e. the sum of counts that would be emitted in the cavity region according to the emissivity profile. The noise $N$ on the other hand is defined as the square root of the sum of counts from the surrounding regions contributing to the projected area that the cavity occupies. En{\ss}lin \& Heinz find that the $S/N$ of an observed bubble depends strongly on their location in the cluster, as well as the specifics of the observations. Using our nomenclature, we generalize their equation (11) for any bubble of size $R_b$ at a radius $r$ as
\begin{equation}\label{e.snbubble}
{S\over N} = { 4R^2_b\, n_0 \left[ 1+(r/r_c)^2 \right]^{-3\beta} \over 3 \left( \left[ 1+\cos(\theta)^2 (r/r_c)^2 \right]^{-3\beta +{1\over2}} +\delta \right)^{1\over 2} }  \left[ { \pi \lambda_T \over r_c\, B(3\beta-{1\over 2}, {1\over 2})} \right]^{1 \over 2},
\end{equation}
where $\theta$ is the angle by which the jet axis is tilted out of the observed sky plane, $\delta$ is the ratio between the peak flux and the background flux (typically $\delta << 1$, but this depends on observational details), $\lambda_T$ is a function of the detector's effective area and the distance of the target, and $B$ is the incomplete beta function. Please refer to \citet{EnsslinBubbledetectability} for further details. 

Figure \ref{f.r_rb_incompleteness} shows how the size of a bubble needs to evolve in order to keep a constant $S/N$ throughout its lifetime, i.e. to be equally likely to be detected. The colored solid lines show the same line, but for different angles $\theta$ with respect to the line of sight. The dotted red lines repeat the line of constant detectability for the face-on case for different initial bubble sizes $R_{b,0}$, spaced by factors of 4. This plot dramatically emphasizes the importance of understanding incompleteness for our sample, as the data points (grey shaded) mostly follow a similar slope as the incompleteness limits. We would like to stress this under-appreciated fact and urge other researchers to take this into account when drawing conclusions about the energy budget of AGN in clusters.

\subsection{Hydrodynamic Instability Effects\label{s.instabilities}}

One of the most astonishing discoveries about cavities in clusters of galaxies has been that these bubbles seem to stay intact over a very long time.  There are a number of hydrodynamical instabilities that should develop in these rising bubbles, namely Kelvin-Helmholtz (KH), Rayleigh-Taylor (RT) and Richtmyer-Meshkov (RM) instabilities.  Here we use simple estimates of the growth-time of these instabilities to both determine whether such instabilities will produce observable features in the cavities and determine which instabilities are most important.

As the bubble rises, the shear flow between the low-density rising bubble and the denser surrounding medium will develop Kelvin-Helmholtz (KH) instabilities. A simple estimate of the growth time for such an instability can be obtained from perturbation theory \citep[e.g.][]{ChandrasekharHydrostability}:
\begin{equation}
T_{\rm KH} \approx {\lambda_h \over \Delta v} {(\rho_{\rm b}+\rho_{\rm
amb})^2 \over \rho_{\rm b}\,\rho_{\rm amb}}
\end{equation}
where $\lambda_h$ is the size scale of the instability and $\Delta v$ is
the velocity of the rising bubble relative to its surrounding medium.
This bubble velocity can be approximated by using several estimates, such as the terminal buoyancy speed, the refill time, or the sound speed in the ambient medium \citep{BirzanCavities}. In \ref{t.bubbleprop}, we decide to list the integrated buoyant rise times according to \citet{EnsslinBubbledetectability} for an adiabatic expansion of the bubbles. 
$\rho_{\rm b}$ and $\rho_{\rm amb}$ are the densities of
the rising bubble and its surroundings respectively.  Typically, values of 
$\rho_{\rm b}<0.1-0.01 \rho_{\rm amb}$ are to be expected from numerical simulation and from limits on the X-ray emission coming from these cavities. In our estimates, we assume a fixed contrast of $0.01$ between the densities, and a rise velocity close to the sound speed. In Table~\ref{t.bubbleprop}, we list the KH growth times for the largest mode that could possibly develop, i.e. for $\lambda_h=R_{\rm b}$, the current size of the bubble. As the density contrast is rather large, the growth of KH instabilities should be rather slow compared to the inferred rise times of the bubbles. Assuming a density contrast of only $0.1$ instead decreases the growth times by a factor of $\sim 8$, which would bring them closer to the vicinity of the bubble ages. Thus, we expect KH instabilities to only play a role on a scale much smaller than the current bubble size, at least $\lambda_h<1/10 \, R_{\rm b}$. While it is unclear what the exact observational signature in X-rays should be, {\it Chandra}'s resolution should in principle already be good enough to find signatures of developing KH instabilities. The fact that the sides of the bubbles seem to be completely intact calls for either a suppression of KH by viscosity of magnetic fields, or a very high density contrast.

\begin{figure}[t]
\begin{center}
\includegraphics[width=0.49\textwidth]{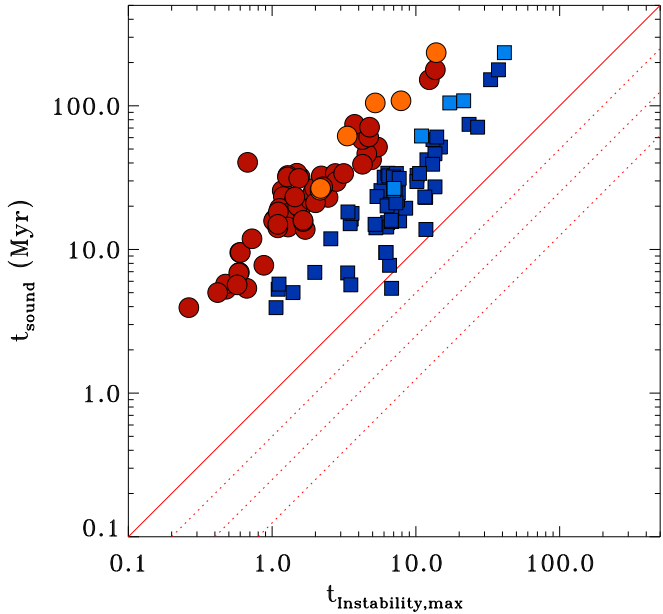}
\end{center}
\caption{Adiabatic rise times for an assumed $\Gamma=4/3$ against the instability time scales for the largest modes ($\lambda_h=R_{\rm b}$) of the Richtmyer-Meshkov (RM, red circles) and Rayleigh-Taylor instabilities (RT, blue squares). The solid line delineates equality between the two timescales. Every bubble above this line should have already developed instabilities on the order of the bubble size. The three dotted lines (top to bottom) show equality smaller modes, namely $\lambda_h/R_{\rm b}= 1/\sqrt{2}$, $1/2$ and $1/\sqrt{8}$ for the RT instability and $\lambda_h/R_{\rm b}= 1/2$, $1/4$ and $1/8$ for the RM instability. \label{f.timescales}}
\end{figure}

The top of the bubble on the other hand is prone to develop Rayleigh-Taylor instabilities, as a heavy fluid (the ambient medium) is accelerated on top of a light fluid (the bubble), with the gravitational acceleration pointing toward the light medium (toward the center of the cluster). Estimates of the growth time for RT instabilities are often given as:
\begin{equation}\label{e.rttimescale}
T_{\rm RT} \approx \left ( \frac{\lambda_h}{2 \pi A |g|} \right )^{0.5} 
\end{equation}
where $g$ is the gravitational acceleration and $A$ is the atwood
number: ($\rho_{\rm amb}-\rho_{\rm b}$)/($\rho_{\rm
b}+\rho_{\rm amb}$).  Because the density contrast
is supposed to be high ($\rho_{\rm b}<0.1-0.01 \rho_{\rm amb}$), the
atwood number\footnote{Note that
the more general estimate for the RT growth time requires knowledge of the entropy gradient at the instability surface. But for our estimates, the simple atwood number prescription provides a safe upper limit on the growth time.} is not larger than about $1.2$. Table~\ref{t.bubbleprop} lists our RT growth time estimate for an atwood number of $1.0$ and the largest possible mode, i.e. the current size of the bubble. The gravitational acceleration has been estimated using the $\beta$ model assuming hydrostatic equilibrium. While these growth time estimates are thus not intended to be used at face value, they show one characteristic that is ubiquitous for the whole sample: they are all on the order of or smaller than the rise time of the bubbles. This is particularly interesting, since we listed the growth time for the largest possible scale length, which is the longest. Thus, RT instabilities should at least have already begun to shred most of the bubbles apart on scales of a significant fraction of the bubble size. In fact, the RT instabilities would have had sufficient time to grow a global instability on the bubble size scale for most of the bubbles (Figure \ref{f.timescales}, blue squares). \citet{SokerRT} have argued that the one should take into account the expansion of the bubble during inflation when computing the RT time scale. This acceleration diminishes the effect of the gravitational acceleration, leading to a suppressed onset of RT instabilities. However, this theoretical work is rather idealized and it has yet to be determined if this mechanism actually works successfully in realistic 3-dimensional hydrodynamic simulations. We have only considered the traditional Rayleigh-Taylor time scales but keep this caveat in mind. 

However, probably the fastest growing instability for the case of our
rising bubbles is the Richtmyer-Meshkov (RM) instability. RM instabilities grow when low-density material (the cavity) ``plows'' through a higher-density ambient medium \citep{Richtmyer, Meshkov}. These instabilities are usually important when the contact surface between two fluids is accelerated (e.g. by a shock passing through). The characteristics of this instability are usually the generation of ``spikes'' separated by voids on the surface. An analytic estimate of the RM instability is given by:
\begin{equation}\label{e.rmtimescale}
T_{\rm RM} \approx {\lambda_h \over 2 \pi A \Delta v}.
\end{equation}
This instability is the fastest growing (at all size-scales) of the three discussed in this section, as shown in Table~\ref{t.bubbleprop}. Again, here we used the bubble size as the largest possible mode, and the sound speed as a proxy for the cavity velocity. Note that the RM growth times are all on the order of Myrs or below, always either on the order of the bubble rise times, or up to an order of magnitude shorter than them (Figure \ref{f.timescales}, red circles). Thus, we conclude that RM instabilities almost certainly should have developed on at least part of the top surface of the bubble.

We urge the reader to remember that these timescales are approximate. It may be that although we predict that instabilities on sizescales equal to the bubble size should develop for most of our cavities, uncertainties in such growth-time estimates prevent us from stating this with complete certainty.  However, the predicted growth time for RM and RT instabilities for 1/10th the bubble size is {\it always} over a factor of 10 lower than the rise time and such instabilities almost certainly should have developed (Figure \ref{f.timescales}). Such instabilities should be observable with the high resolution of {\it Chandra} and are not seen to our knowledge, even in the very deep observations of Hydra~A \citet{WiseHydraA} or Perseus \citep{FabianPerseusripples}. 

RT and RM instabilities should be entirely suppressed for magnetically dominated models for modes larger than the coherence length of the magnetic field, which for the CDJ model is on the order of the bubble size.

Several caveats should be kept in mind when using these simple estimates. First of all, the rise times that we have computed assume that the bubble is inflated at the center. If this assumption is far off, and the bubbles get inflated at radii on the order of the core radius, then the quoted rise time should be considered an upper limit on the rise time. If this is the case however, the jets should be highly supersonic, which would result in the shocks and hot spots at the cavity rims in general. Projection effects on the other hand will cause the rise times to be underestimated. Which of the two effects dominates or whether they cancel each other out on average, is unclear and depends on the details of the jet dynamics.

\section{Monte Carlo Simulations including Incompleteness and Projection}\label{s.montecarlo}

\subsection{Methodology}

To properly address the question of which model can truly reproduce the properties of the cavity sample, one would have to ideally model the complete manual detection process, a very work intensive procedure. We follow a somewhat simpler approach by using a Monte-Carlo (MC) technique that mimics the detection process without the need of human intervention. The outline of the MC process is as follows: 

\begin{enumerate}
\renewcommand{\labelenumi}{(\roman{enumi})}
\setlength{\itemsep}{3pt}
\setlength{\parskip}{0pt}
\setlength{\parsep}{0pt}
\item{Pick the initial bubble size $R_{b,0}$} (i.e. the outburst energy)
\item{Pick the distance $r$ to the center and pick $\phi$ and $\theta$ to randomly orient the bubble with respect to the line of sight}
\item{Determine the probability of detecting the cavity}
\item{Repeat steps (i)--(iii) until you have ``detected'' as many cavities as in the original cavity sample (64)}
\item{Fit the simulated $r$--$R_b$ data set to determine the parameters slope, y-offset and intrinsic width}
\item{Repeat steps (i)--(iv) 1000 times to find the uncertainties in the parameters}
\end{enumerate}
 
The key challenges in this Monte-Carlo bootstrap method are in picking the correct initial bubble size (step i) and in determining whether the evolved bubble would be detected or not (step iii), which are unfortunately also the steps that are the most uncertain. During the rest of this section, we we will follow this basic methodology, but vary the details of these steps. For a valid model, the MC simulations should be able to reproduce the three main characteristics of the distribution of data points in the $r$--$R_b$ plots, namely the slope, y-offset and the intrinsic width, as determined by the fitting method \texttt{bandfit} (see appendix in \citet{DiehlAGN} for a description of this method).

The amount of work that has been done on the detectability of X-ray cavities is very limited. We employ the following criteria to determine whether a bubble is detected or not:
\begin{enumerate}
\renewcommand{\labelenumi}{(\alph{enumi})}
\setlength{\itemsep}{3pt}
 \setlength{\parskip}{0pt}
 \setlength{\parsep}{0pt}
 \item{$S/N$ criterion: We compute the $S/N$ of a bubble from \citet{EnsslinBubbledetectability} using our equation (\ref{e.snbubble}). For simplicity, we then use a Gaussian probability distribution to compute the detection probability.}
 \item{Geometric criterion: If the extent of the bubble overlaps with the center of the cluster, the cavity would be extremely difficult to detect, in particular if there is another bubble on the opposite side of the cluster that overlaps then as well. In practice, we cut every bubble whose radius exceeds the projected distance to the center. }
\item{Hydrodynamic Instability: In the purely hydrodynamic models, we remove all cavities whose rise times exceed the RT or RM growth times according to equations (\ref{e.rttimescale}) and (\ref{e.rmtimescale}). However, tests show that this has very little to no influence on the simulations.}
\item{Detector Size: If the observed distance to the center exceeds half the size of the {\it Chandra} CCD ($4\arcmin$), we remove the candidate from the sample, as most observations in our sample are single-CCD observations.  }

\end{enumerate}
For more details on any of these criteria, please refer back to section \ref{s.discussion}.

Thus, this process is rather complex and depends a large variety of input parameters. One key parameter is the assumed shape of the underlying pressure profile, given by the $\beta$-profile parameters $\beta$ and $r_c$ and $p_0$ (i.e. $n_0$ and ${\rm k}T_0$). The steepness of the pressure gradient is crucial in determining how quickly the bubbles expand while rising. 

Further, observation specific details become important for the detection probability (see section \ref{s.incompleteness} for more details): the effective exposure time $\tau$, the average detector efficiency $\epsilon$, the X-ray background level $S_{\rm bg}$, as well as the distance to the object $d$.  In the following discussion, we will describe our choices for all of these parameters and lay out how these choices influences the behavior of the simulations. 

\subsection{Basic Simulations}

We start with a MC simulation with the most simplistic assumptions possible and demonstrate why this is insufficient to reproduce the observations. The most basic approach one can take to design a replica of a true sample to $0^{\rm th}$ order is to use {\it one} set of fiducial template cluster properties common to all $64$ cavities. In these MC simulations, the morphological and observational parameters of this one template cluster are chosen as the average of the cluster sample, and kept fixed for all cavities. 

It turns out that no model comes even close in reproducing the cavity sample's properties. Slopes are generally too shallow. In fact, the simulated samples behave mostly as expected from the theoretical considerations. The hydrodynamic models AD53 and AD43 and the FML model show the shallowest evolution, but get steepened due to the incompleteness cutoff for small cavities at large radii. The CDJ and CIH models are affected the least by this incompleteness effect, as the evolution is essentially parallel to the incompleteness cutoff, i.e. a bubble that would be detected at small radii is likely to be detected further out as well.  

Since none of the simulations can reproduce the observations, we have to add additional sophistication to our MC simulations. As our sample encompasses a very large variety of environments, pressure profiles, size scales, temperatures, etc., we need to take the intricate interdependence between these parameters into account (e.g. $\beta$ with ${\rm k}T_0$). As we will show in the following sections, the complex correlations between these parameters are the reason this simple test fails. However, we emphasize that the differences in the predictions of the various models are the strongest in this simplified case. Thus, this type of analysis will greatly benefit from having a larger cavity sample in the future, that can be split up into groups with very similar properties. We tested this idea with the current sample, but its size of 64 objects does not allow for sufficient statistical accuracy on the sub-samples to constrain any models. 

\subsection{Use of a Full Set of Parameters}\label{s.mc_data}

\begin{figure*}[t]
\begin{center}
\includegraphics[width=.6\textwidth]{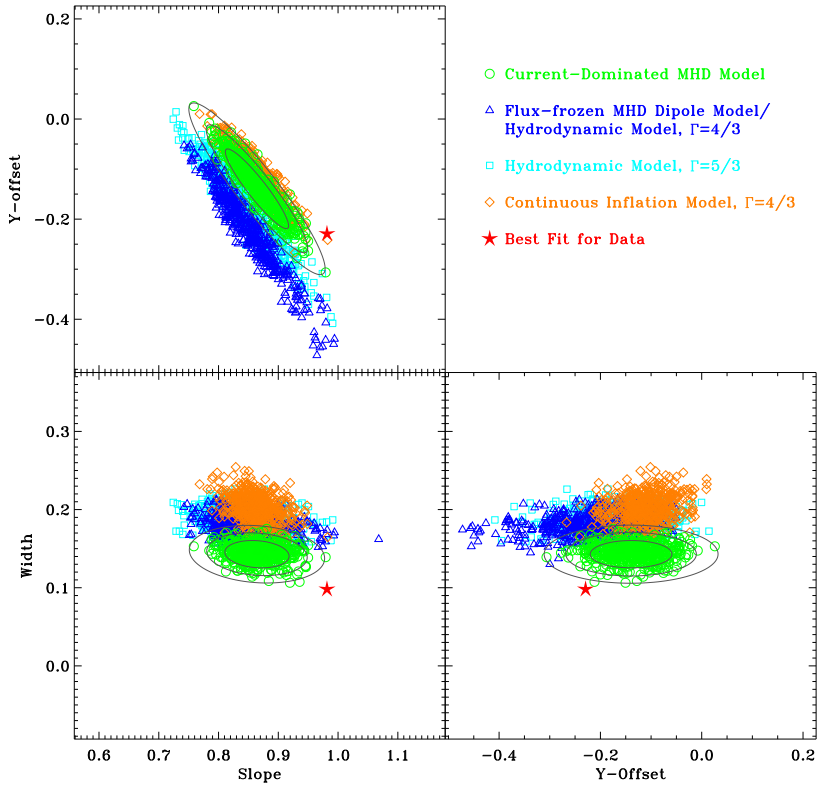}
\end{center}
\caption{Three-parameter plot for one of our Monte-Carlo simulations of section \ref{s.mc_data}. We produced 500 simulations for each model with 64 detected cavities each. The plot shows the best linear fit parameters: slopes, y-offsets and intrinsic widths. In this run, the initial bubble size was derived from the observed bubble sizes. Observational errors and projection effects results in an additional scatter that increases the width of the observed correlation, preventing the simulations to match the observations. Colors indicate the model: CDJ model (green circles); CIH (orange diamonds); FML (dark blue triangles); AD43 (also dark blue triangles) and AD53 (light blue boxes). The red star indicates the location of the best fit through the actual data, that we want to reproduce. The contours show the $1$, $2$ and $3\sigma$ confidence levels for the CDJ model. Note that the CDJ model gets closest to the observed correlation, though still being slightly more than $3\sigma$ away. The shape and size of the shown error ellipse should be also a fair representation of the uncertainty in the fitted parameters themselves (red star).\label{f.mc_data}}
\end{figure*}

In order to improve on the basic simulations, we have to use as much of the cluster properties as possible to reproduce realistically the sample properties. Since we do know the details of the cluster profile and the observations, we use {\it all} of the available parameters to simulate our set of cavities: $\beta$, $r_c$, ${\rm k}T_0$, $n_0$, $p_0$, $d$, $\tau$, $\epsilon$, and $S_{\rm bg}$. Thus, for each cluster in our sample, we can choose an initial bubble size and simulate cavities specifically tailored to each cluster and its observation as well. 

The observed sample itself is essentially a random conglomerate of observations of clusters, groups and galaxies for which cavities have been detected. To recreate this complex selection function, we decide to simulate cavities individually for each object and repeat until the exact number of observed cavities has been matched. 

The advantage of this approach is that it is very tightly constrained, and that we take our host selection function properly into account. The only free parameter in our MC simulations is now the choice of our initial bubble size, and we explore several different prescriptions in how to choose this parameter. First we use the measured cavity sizes to infer back the initial bubble size based on the cavity model that we are investigating at the moment, i.e. we use our theoretical understanding of bubble size evolution (equations (\ref{e.rb_gamma}), (\ref{e.rb_dipole}), and (\ref{e.rb_current})) to calculate the size the cavity would have started with at the center. 

The only problem with this method is that we have to assume a geometry, as we only have the {\it projected} distance to the center available, and that the observational errors in determining the observed bubbles will affect the bubble sizes in our simulations. It turns out that this particular problem makes it impossible to reproduce the full observed sample properties. Figure \ref{f.mc_data} shows the results of the linear fit to our simulated equivalent of Figure \ref{f.r_rb_kpc} ($R_b$ vs. $r$). The plot shows the 3d-parameter space (slope, y-offset and intrinsic width) occupied by our simulated models (light blue: AD53, dark blue: AD43 and FML, orange: CIH, green: CDJ), along with the observed parameters. As one can see, none of the models agree with the data (red star). The current-dominated model comes closest, with the parameter set being just outside the $3\sigma$ contour lines. The main problem with the models in general is that the extremely tight width of the correlation cannot be reproduced. We think that this is the result of the amplifications of the observational errors and the projection effects in the actual sample. In fact the difference between the inferred intrinsic widths of the simulated samples to the observations is close to the value expected from observational error alone (a $10\%$ relative error is $0.4{\rm dex}$). The error in the initial bubble size will not only affect the intrinsic width, but also slope and offset of the correlations. 

The fact that the observed correlation between $r$ and $R_b$ (Figure \ref{f.r_rb_kpc}) is so tight, suggests that the initial bubble size $R_{b,0}$ is intimately connected to other intrinsic cluster properties, which we explore in the next section.

\subsection{Initial Bubble Size $R_{b,0}$ Linked to Core Radius $r_c$}\label{s.mc_rc}

\begin{figure*}[t]
\begin{center}
\includegraphics[width=.6\textwidth]{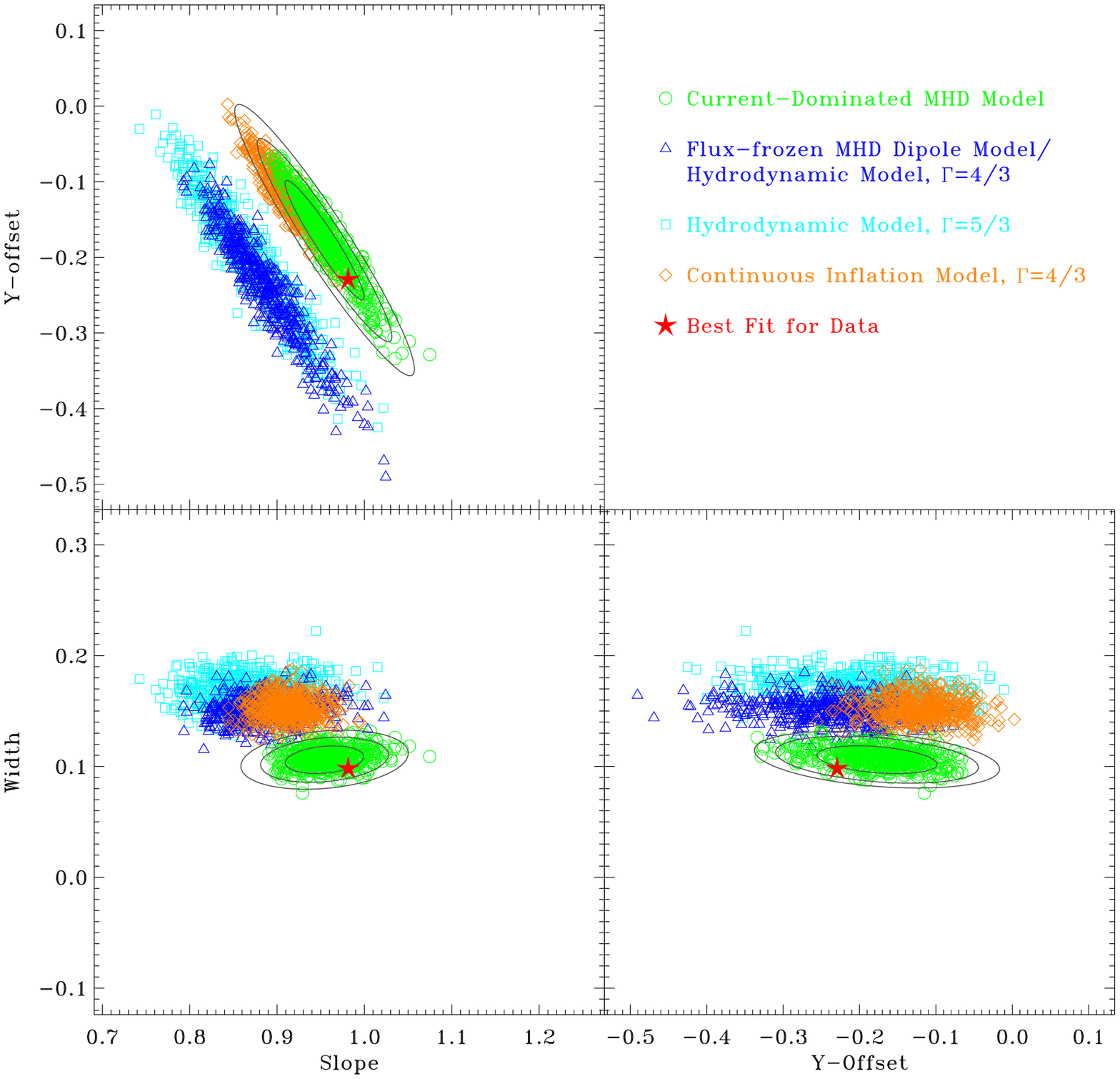}
\end{center}
\caption{Same plot and data symbols as in Figure \ref{f.mc_data}, but this time for the simulation described in section \ref{s.mc_rc}. In this run, the initial bubble size is set to be proportional to the core radius. The contours show the $1$, $2$ and $3\sigma$ confidence levels for the current-dominated model, with the $1\sigma$ contour already encompassing the data parameters in all three plots. Note that all other models can be excluded on more than the $5\sigma$ level. \label{f.mc_rc}}
\end{figure*}

\begin{figure*}[t]
\begin{center}
\includegraphics[width=.2\textwidth]{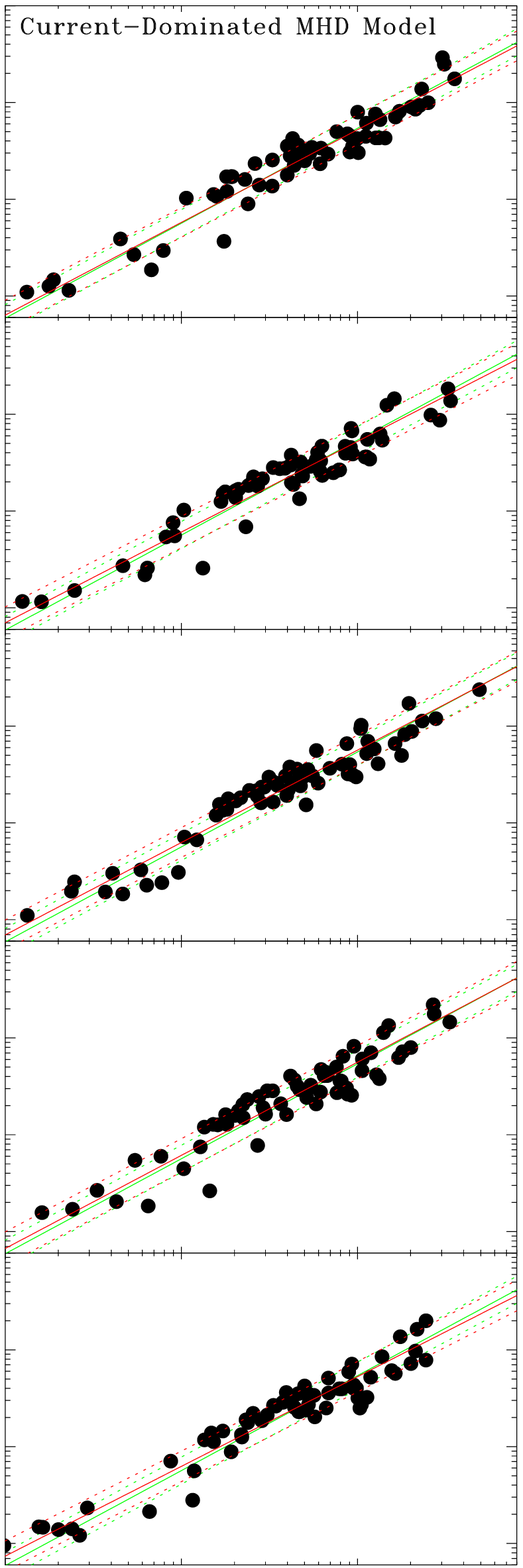}
\hspace{0.02\textwidth}
\includegraphics[width=.2\textwidth]{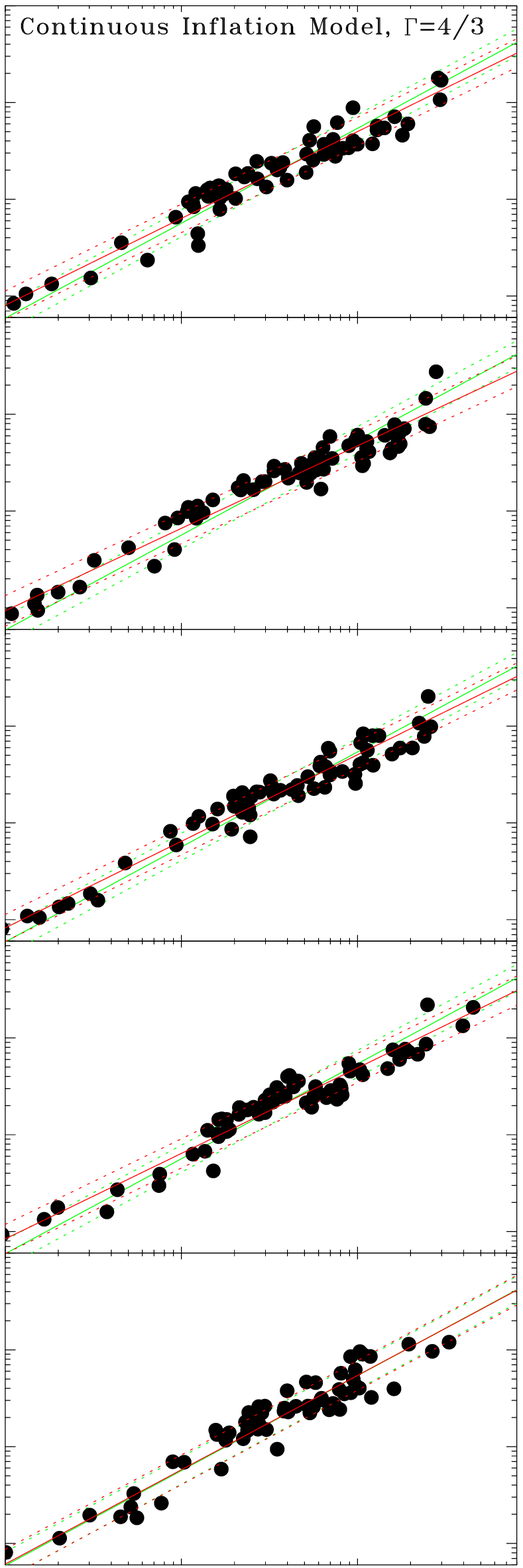}
\hspace{0.02\textwidth}
\includegraphics[width=.2\textwidth]{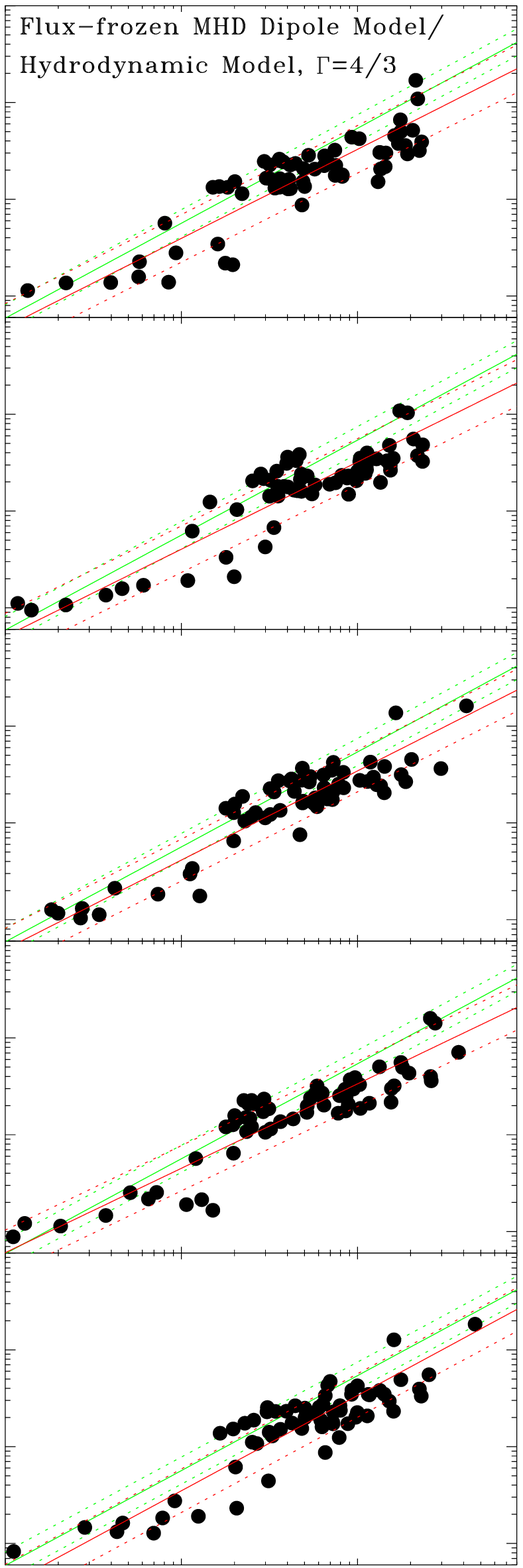}
\hspace{0.02\textwidth}
\includegraphics[width=.2\textwidth]{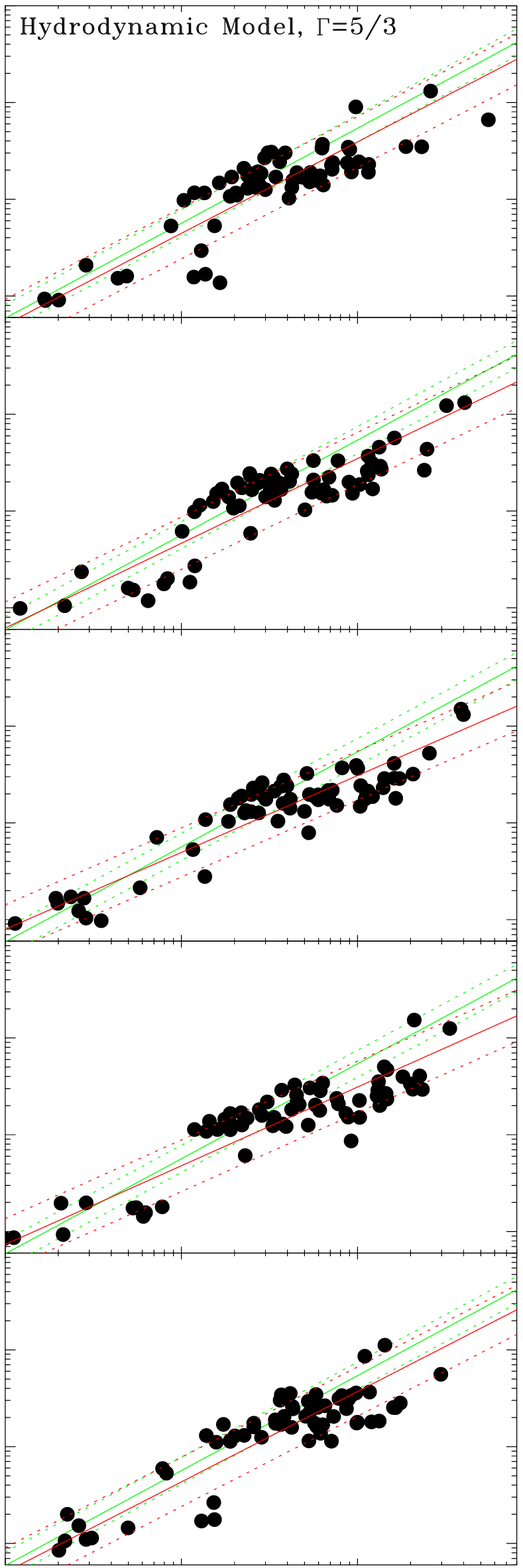}\\
\end{center}
\hspace{.35\textwidth}
\includegraphics[width=0.247\textwidth]{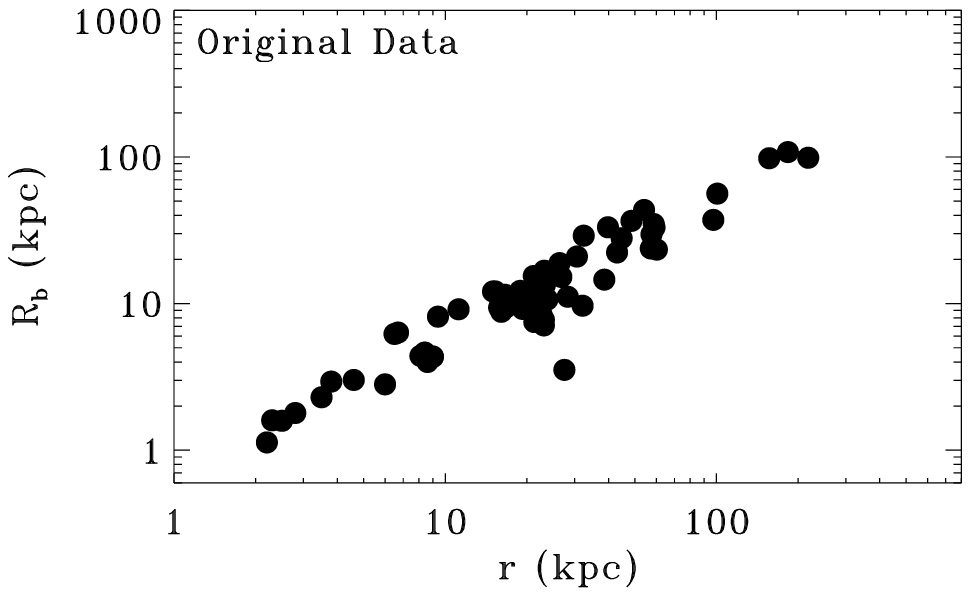}
\caption{Sample plots from our Monte-Carlo runs. The bottom panel shows the original data from our cavity sample. The columns shows 5 of our Monte-Carlo runs for the CDJ, CIH, AD43/FML, and AD53 models (from left to right). The green lines in each plot show the best fit to the actual data, along with the $1\sigma$ intrinsic width. The red lines show the equivalent from the shown simulated data set. Note how the CDJ model reproduces the sample properties best, followed by the CIH model.  \label{f.mcsample}}
\end{figure*}

As the outcome of the last set of MC simulation using the full set of sample parameters has shown promising improvement over the basic simulation, we now look for the best way to set up the initial bubble sizes. Figure set \ref{f.r_rb_mbhrccolor} suggests two possible parameters: the black hole mass $m_{\rm BH}$, and the core radius $r_c$. 

Using the black hole mass as a surrogate for $R_{b,0}$ does not substantially improve the results of the last MC simulation, though it does move the simulation closer to agreement. However, a substantial improvement is achieved when relating $R_{b,0}$ to the core radius. We explore various power-law correlation, but find that a simple proportionality of the type $R_{b,0}={1\over 2} r_c$ works best. Figure \ref{f.mc_rc} shows the parameter space for this MC simuation. Note how only the CDJ model is able to reproduce the data, while all other models can be excluded at more than the $5\sigma$ confidence level. 

We would like to emphasize that even with a large number of combinations of the parameters available to us, we did not manage to bring either the hydrodynamic or the magnetic dipole model into agreement with the observed data, even when only using subsets of the data. In particular, even with fine-tuning,  not a single MC simulation for the AD43, AD53, FML or even CIH models has been able to produce a correlation as tight as the observed correlation. 

Figure \ref{f.mcsample} shows a sample of 5 of our MC simulations for the various models (left to right: CDJ, CIH, FML and AD43, AD53). These samples have been drawn randomly, and not been hand-picked by us. Note that the CDJ samples closely resemble the observed sample (bottom, for reference), followed by the CIH model, while the other two columns do not. They exhibit signs of ``banding'' in the center, where one can actually see the effect of the shallower slope for the evolution. This effect not only makes the fitted slope shallower, but also widens the correlation significantly, which in turn affects the inferred intrinsic width and the Y-offset. 

We conclude that with our current understanding of incompleteness issues, the CDJ model is the only model that we considered that can quantitatively explain the cavity data. Any other correct cavity or jet model would have to mimic the behavior of the CDJ model in terms of the size evolution of cavities. We also explored the possibility of an unknown model whose cavities expand even faster than that of the CDJ model, by simply increasing the exponent $\alpha$. It is interesting to note that those models also fail to reproduce the data. The CDJ model seems to lie in the ``sweet spot'' for the observed tight $r-R_b$ correlation to work. 

Surprisingly, even the CIH model can be formally excluded by the MC simulations, even though its predictions are close to the CDJ model at large radii. The differences at small radii or the slightly shallower slope at large radii ($\sim 17/24$ instead of $\sim 3/4$) are sufficiently to be a worse match. This trend is supported by the relations of $pV$ with $I_z$ and $\dot{E_0}$ in Figure \ref{f.pVIz}, where the CDJ model exhibits a much tighter correlation. Nevertheless, the predictions for the CIH model are sufficiently close to the CDJ model that details in the cavity detection algorithm may yet change this answer. 

\subsection{Caveats}

While the above simulations provide good constraints on the nature of cavities in clusters, we have to be careful in not over-interpreting their results. The MC simulations are based on the assumption that we in fact do understand all the incompleteness issues associated with the data. However, this is not really the case. While \citet{EnsslinBubbledetectability} provide a nice analytic estimate of the $S/N$ of a cavity in a cluster atmosphere, it is unclear how this estimate translates into the real world of actually identifying the cavities. This task is left to experienced astronomers looking at the data individually and detecting bubbles by eye. The only tools guiding them at this task are sometimes radio observations and unsharp-masking. The presence of full or at least partial cool rims around cavities greatly enhances their detectability, which may bias our view of cavities in general. 

There is no automated tool available to detect bubbles. Until this is available, and we can conduct quantitative tests of our ability to detect bubbles, we cannot claim that we truly understand the incompleteness effect for cavities in clusters. 

The fact alone that we have to assume an ad-hoc geometric detection criterion shows how fragile this construct is. Our MC simulations are very sensitive to incompleteness issues, which are essential in producing the tight correlations seen in the data. We suggest that the type of analysis described here should be repeated when incompleteness is better understood and a larger sample is available.

\section{Conclusions and Future Work}\label{s.conclusions}

We present the first systematic analysis of cavity sizes from cavities located in cluster, group and galaxy environments. We summarize predictions from four particular models: three purely hydrodynamic (AD53, AD43 and CIH) and two magnetically dominated models (FML and CDJ, see Table \ref{t.models}). The two simplest hydrodynamic models AD43 and AD53 assume the bubble is formed a priori close to the center and its interior pressure due to gas pressure from a hot, low density gas. The bubbles then adiabatically expand during the buoyant rise in the intracluster medium. These two models differ only in the use of two different adiabatic indeces $\Gamma=5/3$ and $4/3$. The CIH model also assumes adiabatic expansion, but rather continuously injects energy into the cavities. This requires the jet to stay connected to the cavities as they rise, which may be challenging to achieve in reality. 

The first magnetically dominated model (FML) assumes that the magnetic field configuration consists of a dipole magnetic field generated by randomly oriented magnetic flux loops. We show that the predictions from this model are identical to the $\Gamma=4/3$ hydrodynamic model, which makes a distinction of these models from cavity sizes alone unfeasible. Thus, one cannot possibly use cavity sizes to determine the fraction of pressure supplied by magnetic loops threaded through the bubble surface. The last model that we consider is a magnetically dominated model in which the jet carries a current (CDJ). In this model, the bubble size is set by the point where the interior magnetic pressure due to the flowing current is matched by the outside pressure. 

Our analysis of the 64-cavity sample shows that the bubble sizes are much bigger than predicted from the adiabatic and the magnetic dipole model, and that they are more consistent with the continuously inflated hydrodynamic model and the current-dominated jet models instead. In particular we find that the inferred currents in these systems scale almost linearly with the core radius of the cluster and the central black hole mass. 

We conduct Monte-Carlo simulations of the bubble detection process including incompleteness issues, which we also discuss in length. We show that the magnetically dominated model is always favored compared to the other four models. For the case that the initial bubble size is linked linearly to the core radius, the current-dominated model is consistent with the data within one standard deviation. Even with fine-tuning, we are unable to bring the other four models into quantitative agreement with observations, and never get closer than a $\sim 5\sigma$ offset. Any viable future jet model has to at least mimic the behavior of the CDJ model in terms of the size evolution of bubbles.

In order for the purely hydrodynamic model to work, we also need a mechanism to efficiently suppress instabilities. Viscosity \citep{ReynoldsViscousBubble} or magnetic draping \citep{RuszkowskiTangledFields} may provide this added stability. Continuous inflation may suppress the growth of Rayleigh-Taylor instabilities on top of the bubble \citep{SokerRT}. A fine-tuned jet model continuously inflating cavities even at large radii may also result in a faster size evolution of the cavities. However, bringing such a model into agreement with constraints from both X-ray observations and old particle ages inside the cavities will be challenging. 

We stress the fact that we do not consider our results as final. One of our key assumptions is that the bubbles are always in pressure equilibrium with the ambient gas. We base this assumption on the general lack of of shocked gas surrounding the bubbles. If this is incorrect and the bubbles are overpressured closer to the center, they would expand faster than the simple AD53 or AD43 models that we have considered here. To be a bit more quantitative, let us take a fiducial observed bubble of size $R_b$ at $r=2\,r_c$ in a cluster with $\beta=0.5$ and assume that it is currently in pressure equilibrium. If we further take the steeper slope of the CDJ evolution as a given, we can reproject the bubble size back to the center: $R_{b,0}=R_b\,[1+(r/r_c)^2]^{-3/8}\approx 0.55\,R_b$. Since we know its pressure and volume now, as well as its size before, we determine the the bubble's internal pressure $p_{b,0}$ it must have had to start with: $p_{b,0}=p_b\, (V_b/V_{b,0})^{4/3}=p_0\, [1+(r/r_c)^2]^{-3/4} \,(R_b/R_{b,0})^4$, assuming $\Gamma=4/3$. This yields an overpressure of $p_{b,0}/{p_0}\approx 3.34$. Such an overpressure would have driven a shock that should have been easily detectable. A systematic search for such shocks in clusters and galaxies is needed to assess whether all bubbles are overpressured close to the center, but beyond the scope of this paper. However, for the overpressure model to work, we must also explain how all clusters overpressure their bubbles in a very similar fashion to stay in agreement with the correlations discovered in this paper. So far, only Centaurus~A \citep{KraftCentaurusA,KraftCentaurusA2} and NGC~3801 \citep{CrostonShocks} show evidence of shocks surrounding their radio lobes. In addition, an initially overpressured bubble would have adjusted its bubble size on the order of a sound crossing time for the bubble, which is much shorter than the observed bubble ages.

We also do emphasize that the current assessment of incompleteness effects may be oversimplified and influence these results. In order to properly address this important issue, an automated bubble detection tool is desperately needed. Incompleteness has a considerable impact on the energy budget associated with AGN feedback, and has to be taken into account in further studies of the subject. We also suggest that the current analysis be repeated with a larger and more homogenous sample, as the predictions of the models will be much easier to distinguish from each other. 

We provide a framework to test not only currently available models, but that can easily be extended to future models of jet-lobe systems. We encourage modelers to find theoretical considerations predicting cavity sizes as a function of distance to cluster centers, or to extract this information directly from their numerical simulations. 

\acknowledgments

We thank Wise et al. for allowing us to reproduce their Hydra~A figure, and Brian McNamara and Masanori Nakamura for stimulating discussion on the subject. This work was carried out in part under the auspices of the National Nuclear Security Administration of the Department of Energy (DOE) at Los Alamos National Laboratory and supported by a LANL/LDRD program.

\bibliographystyle{apj}
\bibliography{../bibtex/allreferences.bib}

\end{document}